\documentclass[twoside,twocolumn,9pt]{article}
\usepackage{extsizes}
\usepackage[super,sort&compress,comma]{natbib} 
\usepackage[version=3]{mhchem}
\usepackage[left=1.5cm, right=1.5cm, top=1.785cm, bottom=2.0cm]{geometry}
\usepackage{balance}
\usepackage{mathptmx}
\usepackage{sectsty}
\usepackage{graphicx} 
\usepackage{bm}
\usepackage{lastpage}
\usepackage[format=plain,justification=justified,singlelinecheck=false,font={stretch=1.125,small,sf},labelfont=bf,labelsep=space]{caption}
\usepackage{float}
\usepackage{fancyhdr}
\usepackage{fnpos}
\usepackage[english]{babel}
\addto{\captionsenglish}{%
  
}
\usepackage{array}
\usepackage{droidsans}
\usepackage{charter}
\usepackage[T1]{fontenc}
\usepackage[usenames,dvipsnames]{xcolor}
\usepackage{setspace}
\usepackage[compact]{titlesec}
\usepackage{hyperref}

\usepackage{latexsym}
\usepackage{amsmath}
\usepackage{amsfonts}
\usepackage{amssymb}
\usepackage{amsbsy}
\usepackage{wasysym}
\usepackage{marvosym}
\usepackage{oplotsymbl}

\newcommand{\cblue }{\color{black}}

\usepackage{epstopdf}%This line makes .eps figures into .pdf - please comment out if not required.

\definecolor{cream}{RGB}{222,217,201}

\begin{document}

\pagestyle{fancy}
\thispagestyle{plain}
\fancypagestyle{plain}{
%%%HEADER%%%
\renewcommand{\headrulewidth}{0pt}
}
%%%END OF HEADER%%%

%%%PAGE SETUP - Please do not change any commands within this section%%%
\makeFNbottom
\makeatletter
\renewcommand\LARGE{\@setfontsize\LARGE{15pt}{17}}
\renewcommand\Large{\@setfontsize\Large{12pt}{14}}
\renewcommand\large{\@setfontsize\large{10pt}{12}}
\renewcommand\footnotesize{\@setfontsize\footnotesize{7pt}{10}}
\makeatother

\renewcommand{\thefootnote}{\fnsymbol{footnote}}
\renewcommand\footnoterule{\vspace*{1pt}% 
\color{cream}\hrule width 3.5in height 0.4pt \color{black}\vspace*{5pt}} 
\setcounter{secnumdepth}{5}

\makeatletter 
\renewcommand\@biblabel[1]{#1}            
\renewcommand\@makefntext[1]% 
{\noindent\makebox[0pt][r]{\@thefnmark\,}#1}
\makeatother 
\renewcommand{\figurename}{\small{Fig.}~}
\sectionfont{\sffamily\Large}
\subsectionfont{\normalsize}
\subsubsectionfont{\bf}
\setstretch{1.125} %In particular, please do not alter this line.
\setlength{\skip\footins}{0.8cm}
\setlength{\footnotesep}{0.25cm}
\setlength{\jot}{10pt}
\titlespacing*{\section}{0pt}{4pt}{4pt}
\titlespacing*{\subsection}{0pt}{15pt}{1pt}
%%%END OF PAGE SETUP%%%

%%%FOOTER%%%
\fancyfoot{}
\fancyfoot[LO,RE]{\vspace{-7.1pt}\includegraphics[height=9pt]{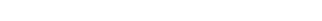}}
\fancyfoot[CO]{\vspace{-7.1pt}\hspace{13.2cm}\includegraphics{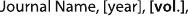}}
\fancyfoot[CE]{\vspace{-7.2pt}\hspace{-14.2cm}\includegraphics{Results/RF}}
\fancyfoot[RO]{\footnotesize{\sffamily{1--\pageref{LastPage} ~\textbar  \hspace{2pt}\thepage}}}
\fancyfoot[LE]{\footnotesize{\sffamily{\thepage~\textbar\hspace{3.45cm} 1--\pageref{LastPage}}}}
\fancyhead{}
\renewcommand{\headrulewidth}{0pt} 
\renewcommand{\footrulewidth}{0pt}
\setlength{\arrayrulewidth}{1pt}
\setlength{\columnsep}{6.5mm}
\setlength\bibsep{1pt}
%%%END OF FOOTER%%%

%%%FIGURE SETUP - please do not change any commands within this section%%%
\makeatletter 
\newlength{\figrulesep} 
\setlength{\figrulesep}{0.5\textfloatsep} 

\newcommand{\topfigrule}{\vspace*{-1pt}% 
\noindent{\color{cream}\rule[-\figrulesep]{\columnwidth}{1.5pt}} }

\newcommand{\botfigrule}{\vspace*{-2pt}% 
\noindent{\color{cream}\rule[\figrulesep]{\columnwidth}{1.5pt}} }

\newcommand{\dblfigrule}{\vspace*{-1pt}% 
\noindent{\color{cream}\rule[-\figrulesep]{\textwidth}{1.5pt}} }

\makeatother
%%%END OF FIGURE SETUP%%%

%%%TITLE, AUTHORS AND ABSTRACT%%%
\twocolumn[
  \begin{@twocolumnfalse}
{\includegraphics[height=30pt]{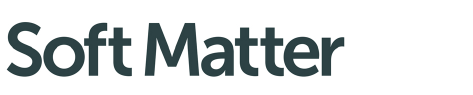}\hfill\raisebox{0pt}[0pt][0pt]{\includegraphics[height=55pt]{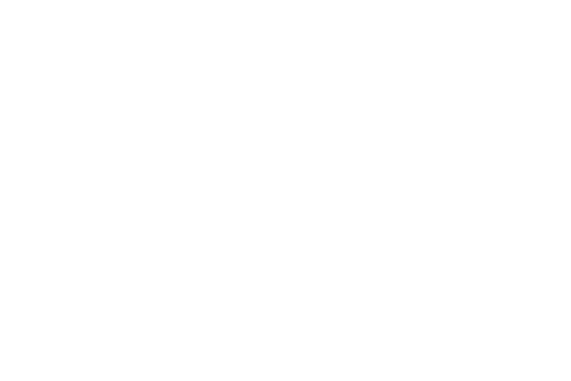}}\\[1ex]
\includegraphics[width=18.5cm]{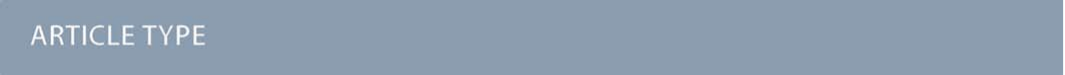}}\par
\vspace{1em}
\sffamily
\begin{tabular}{m{4.5cm} p{13.5cm} }

\includegraphics{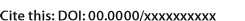} & \noindent\LARGE{\textbf{Collective dynamics of active dumbbells near circular obstacle$^\dag$}} \\%Article title goes here instead of the text "This is the title"
\vspace{0.3cm} & \vspace{0.3cm} \\

 & \noindent\large{Chandranshu Tiwari$^{\ast}$\textit{$^{a}$}, and Sunil P. Singh\textit{$^{b}$}} \\%Author names go here instead of "Full name", etc.

\includegraphics{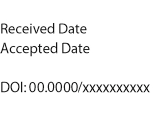} &
\noindent\normalsize{In this article, we present the collective dynamics of active dumbbells in the presence of a static circular obstacle using Brownian dynamics simulation. The active dumbbells aggregate on the surface of a circular obstacle beyond a critical radius ($R_o^c \approx 10$). The aggregation is non-uniform along the circumference, and the aggregate size increases with the activity $(Pe)$ and the curvature radius ($R_o$). {\cblue The dense aggregate of active dumbbells displays persistent rotational motion with a certain angular speed, which linearly increases with the activity.} Further, we show the strong polar ordering of the active dumbbells within the aggregate.  The polar ordering exhibits a long-range correlation, with the correlation length corresponding to the aggregate size.  Additionally, we show that the residence time of an active dumbbell on the obstacle surface grows rapidly with area fraction due to many-body interactions that lead to a slowdown of the rotational diffusion.  The article further considers the dynamical behavior of a tracer particle in the solution of active dumbbells. Interestingly, the speed of the passive tracer particle displays a crossover from monotonically decreasing to increasing with the tracer particle's size upon increasing the dumbbells' speed. {\cblue Furthermore, the effective diffusion of the tracer particle displays the non-monotonic behavior with area fraction; the initial increase of the diffusivity is followed by a decrease for larger area fraction. } }

\\%The abstract goes here instead of the text "The abstract should be..."

\end{tabular}

 \end{@twocolumnfalse} \vspace{0.6cm}
]
  
%%%END OF TITLE, AUTHORS AND ABSTRACT%%%

%%%FONT SETUP - please do not change any commands within this section
\renewcommand*\rmdefault{bch}\normalfont\upshape
\rmfamily
\section*{}
\vspace{-1cm}

%%%FOOTNOTES%%%

\footnotetext{\textit{$^{a}$~Department of Physics, Indian Institute Of Science Education and Research, Bhopal 462 066, Madhya Pradesh, India; E-mail: chandranshu21@iiserb.ac.in }}
\footnotetext{\textit{$^{b}$~Department of Physics, Indian Institute Of Science Education and Research, Bhopal 462 066, Madhya Pradesh, India; E-mail: spsingh@iiserb.ac.in }}

%Please use \dag to cite the ESI in the main text of the article.
%If you article does not have ESI please remove the the \dag symbol from the title and the footnotetext below.
%\footnotetext{\dag~Electronic Supplementary Information (ESI) available:[details of any supplementary information available should be included here]. See DOI: 00.0000/00000000.}
%additional addresses can be cited as above using the lower-case letters, c, d, e... If all authors are from the same address, no letter is required

%\footnotetext{\ddag~Additional footnotes to the title and authors can be included \textit{e.g.}\ `Present address:' or `These authors contributed equally to this work' as above using the symbols: \ddag, \textsection, and \P. Please place the appropriate symbol next to the author's name and include a \texttt{\textbackslash footnotetext} entry in the correct place in the list.}

\section{Introduction}

% \textit{Introduction\textemdash}
The collection of micro-swimmers such as bacteria, algae, amoeba, sperms,  etc., exhibit rich collective dynamics in bulk, and the proximity of solid interface\cite{elgeti2016microswimmers,poon2013clarkia,dombrowski2004self,ishimoto2018hydrodynamic,kumar2019trapping,marchetti2013hydrodynamics,di2010bacterial,cates2012diffusive,ramaswamy2010mechanics,elgeti2015physics,zottl2016emergent,kuhr2019collective,zhang2010collective}. Notably,  phase separation at low density\cite{liu2019self,fily2014dynamics}, circular motion, spontaneous vortex formation in microfluidic devices\cite{wioland2013confinement,lushi2014fluid,suma2014motility}, bacterial turbulence at low Reynolds number\cite{lauga2009hydrodynamics,peng2021imaging,dunkel2013fluid}, giant density fluctuations\cite{marchetti2013hydrodynamics,zhang2010collective},   formation of a periodic array of vortices of bacterial colonies\cite{nishiguchi2018engineering}, self-organization of unicellular organisms in fruiting bodies\cite{liu2019self}, biofilm formation, etc\cite{hall2004bacterial,keller2019study}. Dynamics of such microswimmers in confined environments have attracted immense research interest due to plenty of non-intriguing physical behavior observed, viz., the propensity to swim upstream, helical or chaotic trajectories of motion, tumbling motion\cite{rothschild1963non,woolley2003motility,kaya2012direct,lauga2006swimming,hill2007hydrodynamic,si2020self}, rectification of swimming speed, and its aggregation\cite{di2010bacterial}. Furthermore, numerous experiments have revealed aggregation of micro-swimmers on the solid interface that leads to segregation of slow-moving ordered phase (near surface) and fast-moving random phase (bulk); such aggregate also displays intriguing rotational motion\cite{wioland2013confinement,lushi2014fluid,nishiguchi2018engineering,dunkel2013fluid}.

 Some of the characteristic features of such a complex system are successfully captured in the simple analytical and simulation models\cite{stenhammar2013continuum,ramaswamy2010mechanics,elgeti2016microswimmers,fily2012athermal,redner2013structure,bialke2013microscopic,cates2013active,cates2015motility,suma2014motility,suma2014dynamics,das2020morphological,anand2019behavior,wu2000particle,angelani2011effective,dolai2018phase,jose2021phase,yadav2023dynamics,kumar2023dynamics,bhattacherjee2019re,sese2021phase}. The ellipsoid-like, elongated particles are often treated as a model system for {\it bacterial suspensions} in the literature for the study of the motility-induced phase separation (MIPS)\cite{suma2014dynamics,baskar_quorum,}, aggregation on the solid interface\cite{elgeti2016microswimmers,bar2020self},  collective dynamics induced by steric interactions,  vortex formation in confined environment\cite{wioland2013confinement,lushi2014fluid}, the role of fluid flow \cite{anand2019behavior,zottl2013periodic,zottl2014hydrodynamics,anand2021migration}, etc.  The active dumbbell particle can be entrapped around static circular obstacles by the orbital motion; the mechanism behind entrapment may be hydrodynamic in nature.  \cite{chopra2022geometric,spagnolie2015geometric,sipos2015hydrodynamic,takagi2014hydrodynamic}. Additionally, the large aggregate of spherical particles in certain parameter space spontaneously forms a vortex without any symmetry-breaking facets in the system\cite{pan2020vortex,mokhtari2017collective}.  This emergent phenomenon purely appears from the bias of accepting the incoming particles around the rotating clusters\cite{pan2020vortex,mokhtari2017collective}. On the other hand, anisotropic-shaped self-propelled particles confined in the 2D geometry show interesting phenomena, such as the translation motion of the hedgehog structure and vortex formation of {\it Bacillus subtilis}  bacterial suspension\cite{deblais2018boundaries,wioland2013confinement}. Despite their relevance in understanding microbial dynamics in microfabricated structures and porous media,
  the characteristic features of collective dynamics of the elongated-shaped micro-swimmers in proximity to convex-shape static obstacles are still lacking\cite{wioland2013confinement,pan2020vortex,costanzo2012transport}.

{\cblue Our work aims to present a comprehensive study of the collective dynamics of anisotropic-shaped active particles in the presence of a convex-shaped static obstacle in the two-dimension by exploiting over-damped Brownian dynamics simulations. The anisotropic-shaped active particles are modeled by connecting two disk-like monomers via a spring potential called active dumbbells. Furthermore, our work aims to provide the transport behavior of the tracer particle in the solution of active dumbbells. This can be qualitatively compared to the experimental work of the tracer particle's dynamics in the bacterial bath\cite{patteson2016particle,mino2011enhanced,jepson2013enhanced,leptos2009dynamics}.

We demonstrate that the active dumbbells aggregate on the surface of the static obstacle only if the radius of the obstacle $R_o\geq10$. The aggregation is non-uniform along the circumference and grows with the speed of the dumbbells and the obstacle's curvature radius $R_o$. Remarkably, these dense aggregates of active dumbbells exhibit fascinating collective dynamics on the static obstacle, displaying persistent rotational motion.  The rotational motion can be clockwise or counter-clockwise; such motion is stable for a long time and appears spontaneously.  Furthermore, our simulations demonstrate strong polar ordering of active dumbbells within the aggregate,  where they align themselves on the surface in a spiral form. } 

The active dumbbells align on the surface at a certain angle, which is normal to the obstacle's surface. This angle decreases with the curvature radius and increases with the speed of the dumbbells. The polar ordering of the active dumbbells on the surface displays long-range ordering with the correlation length corresponding to the aggregate size. We elucidate the physical mechanism underlying the aggregation of dumbbells on the convex surface by computing the residence time. The residence time increases rapidly with the concentration of active dumbbells due to many-body interactions.

{\cblue In another scenario, we consider a tracer particle free to move in response to the active force arising from the collision with the active dumbbells in the medium.  Our findings indicate that the directed speed of larger-sized tracer particle surpasses that of smaller-sized ones and displays non-monotonic behavior. Furthermore, the effective diffusion of the tracer particle also displays the non-monotonic behavior with area fraction. The initial increase of the diffusivity is followed by a decrease upon increasing the area fraction, which is very different from the passive solution, where it monotonically decreases with the fraction.  } 

%In such situations, we found that the average speed of the tracer particle decreases with the obstacle's size in the limit of low speed of dumbbells; however, it increases before approaching a plateau value beyond the critical speed of dumbbells.  

%we have shown that the directed speed of bigger-sized tracer particles moves faster than smaller-sized ones.
%We observe that the estimated average speed of tracer particles decreases, particularly in the limit of small self-propulsion speed,  with the size of the obstacle. However, beyond a critical speed, it exhibits an increase before eventually approaching to a plateau value.

The structure of the article is as follows: Section 2 discusses the simulation model of the active dumbbells and simulation parameters.  The simulation results are given in Section 3, which further consists of four subsections. These subsections provide detailed insights into aggregation,  orientational ordering, dynamics of active dumbbells, and dynamics of passive tracer particle within the active medium. The discussion and summary of the results are presented in the conclusion, Section 4.

\section{Simulation Model}
% \textit{Model\textemdash} 
We model $N_d$  polar active dumbbells representing anisotropic-shaped microswimmers in the two-dimensional box with periodic boundary conditions.  Additionally, a disc-shaped static circular obstacle of radius $R_o$ is embedded at the center of the simulation box. The active dumbbells  comprise  two identical monomers of diameter $\sigma$ connected by the harmonic potential $U_s$:
\begin{equation}
    U_s = \sum_{i=1}^{N_d}\frac{k_s}{2}(|\mathbf{r}_{2i} - \mathbf{r}_{2i-1}| - l_0)^2,
    \label{eq:1}
\end{equation}
where $\mathbf{r}_{2i}$ and $\mathbf{r}_{2i-1}$ are the position vectors  of monomers of the ${i}^{th}$ dumbbell, $l_0$ is the equilibrium bond length, and $k_s$ is the spring constant, here $i$ varies from $1,2,....N_d$. To avoid the overlap among various dumbbells, the excluded volume interaction is employed by truncated repulsive Lennard-Jones potential (LJ), 
\begin{equation}
U_{LJ} = \sum_{i = 1}^{2N_d - 1}\sum_{j = i+1}^{2N_d} 4\epsilon \left[\left(\frac{\sigma}{r_{ij}}\right)^{12} - \left(\frac{\sigma}{r_{ij}}\right)^{6}  + \frac{1}{4} \right]. 
\label{Eq:LJ}
\end{equation}
The LJ potential for a given pair $U_{LJ}=0$ for $r_{ij} \ge  2^{1/6}\sigma$, else given by above Eq.\ref{Eq:LJ}, where $r_{ij} = |\mathbf{r}_{i} - \mathbf{r}_{j}|$ is the distance between a pair $i$ and $j$, and $\epsilon$ is strength of LJ repulsion. 

The interaction of monomers with the static circular obstacle is also employed via short-range LJ potential given in Eq.~\ref{Eq:LJ} if the distance between the monomer and the obstacle's wall is less than $2^{1/6}(\sigma/ 2)$ else they do not feel force from the obstacle. 
 %The repulsive LJ potential is employed reached by considering a hypothetical monomer on the periphery of the obstacle. When any dumbbell approaches the obstacle, the monomer interacts with this hypothetical monomer via the same WCA potential given in equation \ref{eq:2}. \\

%$k_B T$ is the thermal energy (with $k_B$ is Boltzmann constant and $T$ is temperature).

The self-propulsion force $f_a$  is imposed along the direction of the bond vector connecting tail to head monomers\cite{suma2014dynamics,suma2014motility}.  Therefore active force on the $i^{th}$ dumbbell is given as $\mathbf{F}_a^i = 2f_a \hat{\mathbf{n}_i}$, here $f_a$ is the strength of active force and $\hat{\mathbf{n}}_i$ is the unit vector given by $\hat{\mathbf{n}}_i = \frac{\mathbf{r}_{2i} - \mathbf{r}_{2i-1}}{|\mathbf{r}_{2i} - \mathbf{r}_{2i-1}|}$.

The overdamped Langevin equation governs the equation of motion of a monomer,
\begin{equation}
 \frac{\partial{\mathbf{r}_{i}}}{\partial{t}} = \frac{1}{{\gamma}_t} \left[- \mathbf{\nabla}_i  U_{LJ}  - \mathbf{\nabla}_{i}  U_s + \mathbf{F}_w^i + f_a \hat{\mathbf{n}}_i  \right] + \mathbf{\eta}_{i}^T,
 \label{Eq:motion}
\end{equation}
where  $\gamma_t$ is the viscous drag, ${\eta_i}^T$ is Gaussian white noise having zero mean, and  its correlation is expressed in terms of the translational diffusion coefficient $D_T$ of a monomer, 
 $ \langle {{\bf \eta}_{i}}^T(t) \cdot {{\bf\eta}_{j}}^T(t^{\prime}) \rangle = 4 D_T  {\delta}_{i j} \delta(t-t^{\prime})$, in two dimension,  and $\mathbf{F}_w^i$ is the interaction with the static obstacle. {\cblue  The solvent-mediated long-range hydrodynamic interactions are neglected for simplicity of the model. } 
 
\subsection{ Simulation Parameters} All the physical parameters are expressed in units of LJ diameter $\sigma$, the thermal energy  $k_BT$, and 
 the translational diffusion coefficient $D_T$. The simulation time is in units of $\tau=\sigma^2/D_T$. The simulation parameters are chosen here as spring constant in the range of $k_s = 10000k_BT/l^2_0$ to $50000k_BT/l^2_0$, $l_0/\sigma= 1$, $\epsilon/{k_BT} = 1$, and length of simulation box  $L/\sigma=150$.  Further, we express the strength of the active force as a dimensionless quantity called P\'eclet number $Pe$, defined as the ratio of active force with thermal force $Pe = f_a l_0/ (k_B T)$.  { \cblue The P\'eclet number $Pe$ can be equivalently referred to as the directed speed of an active dumbbell; they are related as $v_d=2Pe D^0_d/\sigma$.}   The area fraction of the dumbbells is given by $\phi = \frac{N_d\pi\sigma^2}{8[L^2 - \pi R_o^2]}$, here $N_d$ is the number of dumbbells, and  $R_o$ is the obstacle radius.  The area fraction of the dumbbells is kept fixed throughout the manuscript at $\phi=0.2$ unless specified in the plots.  The obstacle radius is varied in the range of  $2.5 -25$ while keeping $\phi$ and $L$ fixed.  To maintain the same packing fraction of the self-propelled dumbbells for various $R_o$, we vary the number of dumbbells from $2500$ to $3000$. The  P\'eclet number $Pe$ is varied from $0$ to $150$. The equations of motion (Eq.~\ref{Eq:motion}) are integrated by the Euler algorithm with Gaussian distributed random displacements,\cite{ermak1978brownian} with steps ranging from $10^{-4} \tau$ to $2 \times 10^{-6} \tau$. Each simulation point is averaged over at least 10 independent simulation samples.

 %We have also performed simulations with a bigger system size, which includes a number of 6000 dumbbells. The results obtained in those simulations are qualitatively similar to the work in the manuscript. 

{\cblue Our simulation parameters and their considered range can be easily compared with experimental scales\cite{chopra2022geometric,spagnolie2015geometric,takagi2014hydrodynamic}. Typically, the size of the bacteria and artificial micro-swimmers falls in the range of $1-10\mu m$, and their swimming speed lies in the range of   2$\mu m/s$ to 28$\mu m/s$\cite{lushi2014fluid,wioland2013confinement,chopra2022geometric,krishnamurthi2022interactions,bhattacharjee2019bacterial}.   Considering the size of the active dumbbell in our simulations is $2\mu m$ at room temperature in the fluid viscosity of $\eta_s=10^{-3}Pa.s$, we can estimate the swim speed using the expression of the P\'eclet number given as $v_d = 2 Pe D_d^0/\sigma$, where $D_d^0$ is the diffusion coefficient of the active dumbbell.  This expression provides the active dumbbell's speed in the range of $0\mu m/s-34\mu m/s$ for the considered simulation range of activity $0-150$ (Pe).  The obstacle or the micropillar in the previous experimental works are considered nearly of $ R_o = 10-20\mu m $\cite{chopra2022geometric,krishnamurthi2022interactions,spagnolie2015geometric}.  This further makes our simulation window of the aspect ratio $ R_o/R_d $ ranging from 2.5 to 12.5, which falls in the range of the experiment scale $2-8$\cite{chopra2022geometric,wioland2013confinement,lushi2014fluid,krishnamurthi2022interactions,spagnolie2015geometric}. }

\section{Results}
%All simulations are performed here at low packing fraction  $\phi = 0.2$, .i.e, below the threshold packing of motility-induced phase separation for all $Pe$. 
The active dumbbells aggregate on the solid interfaces even below the threshold packing of motility-induced phase separation\cite{suma2014motility,suma2014dynamics}; the static interface and self-propulsion speed nucleate the aggregation. Their aggregation propensity becomes prominent for higher swimming speed, larger curvature radius, and hydrodynamics interactions, which may further enhance wall-induced attractions \cite{sipos2015hydrodynamic,anand2019behavior}. Here, we systematically investigate the aggregation behavior of active dumbbells, their orientational ordering and correlations, dynamical behavior on the surface, and the transport behavior of the tracer object as a specific case in the solution of active dumbbells.

\begin{figure}[t]
    \centering
    \includegraphics[width=0.49\columnwidth]{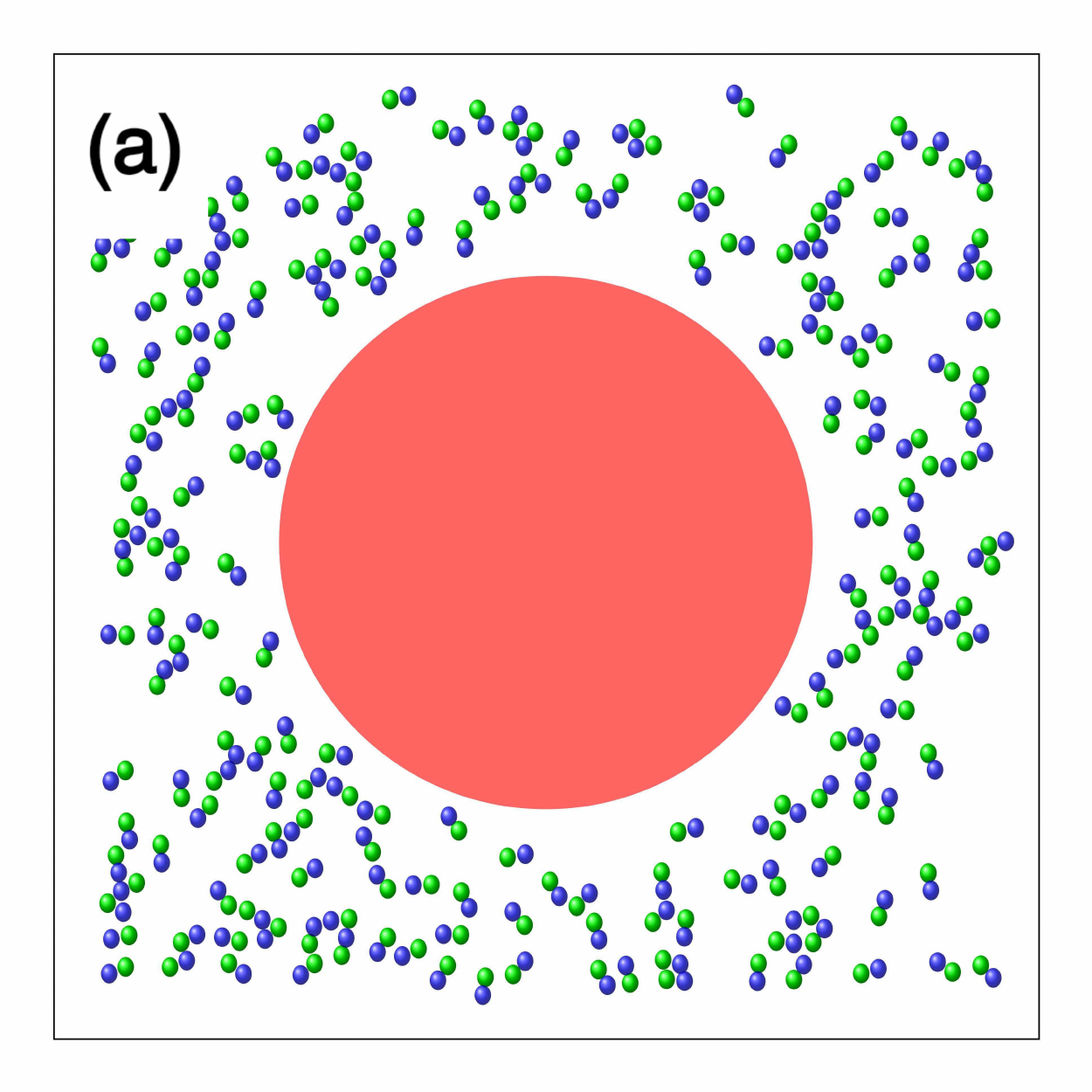}
     \includegraphics[width=0.49\columnwidth]{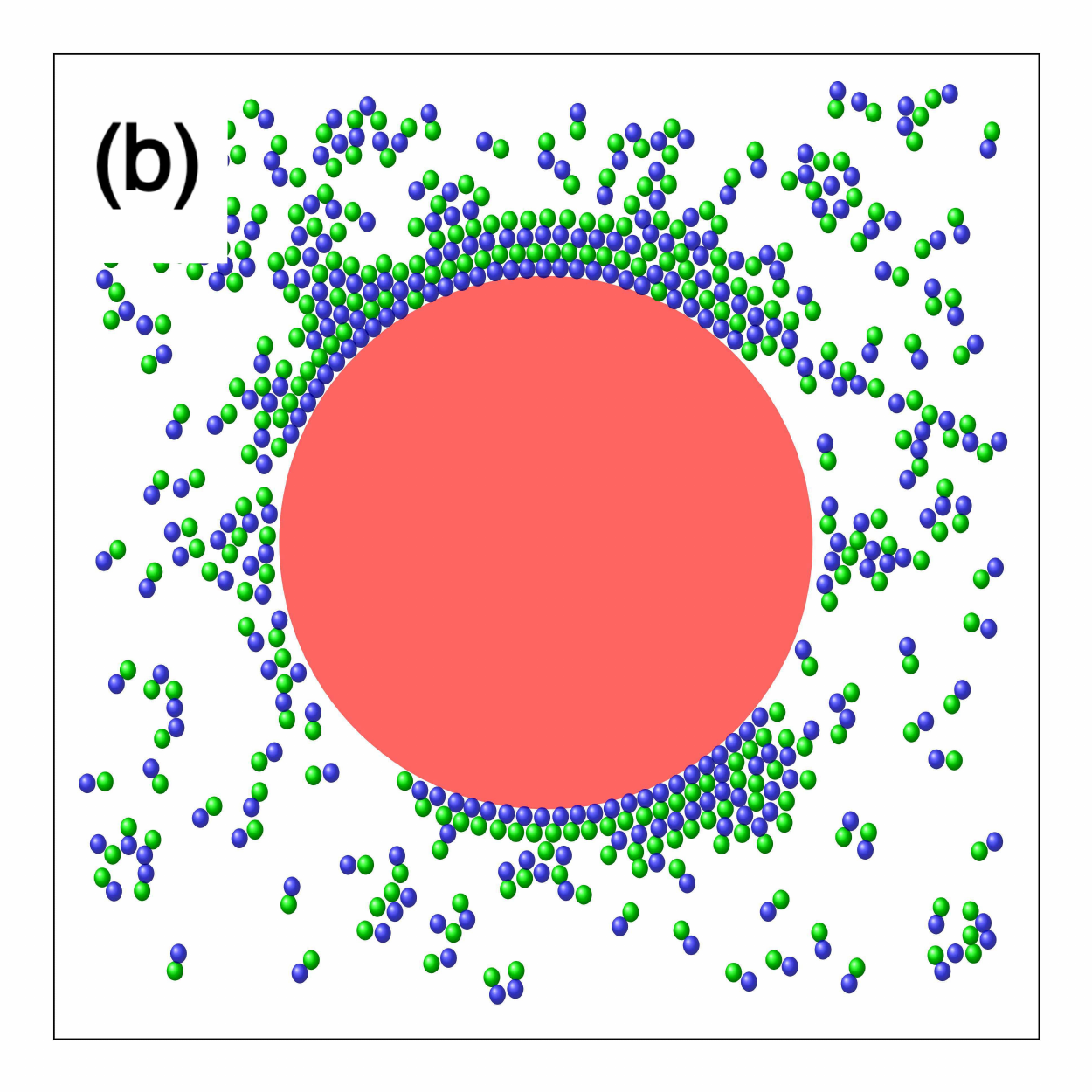}
     \includegraphics[width=0.49\columnwidth]{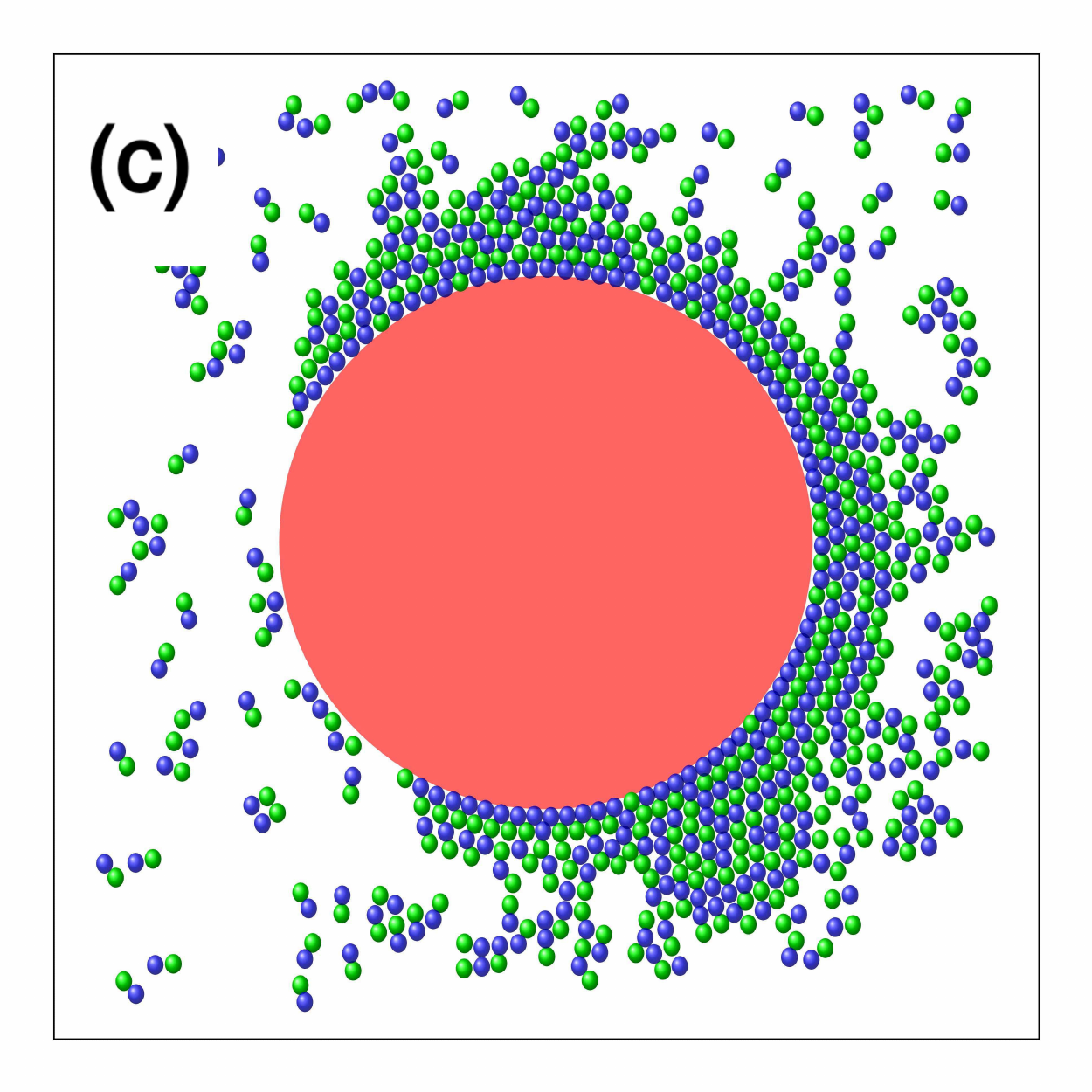}
     \includegraphics[width=0.49\columnwidth]{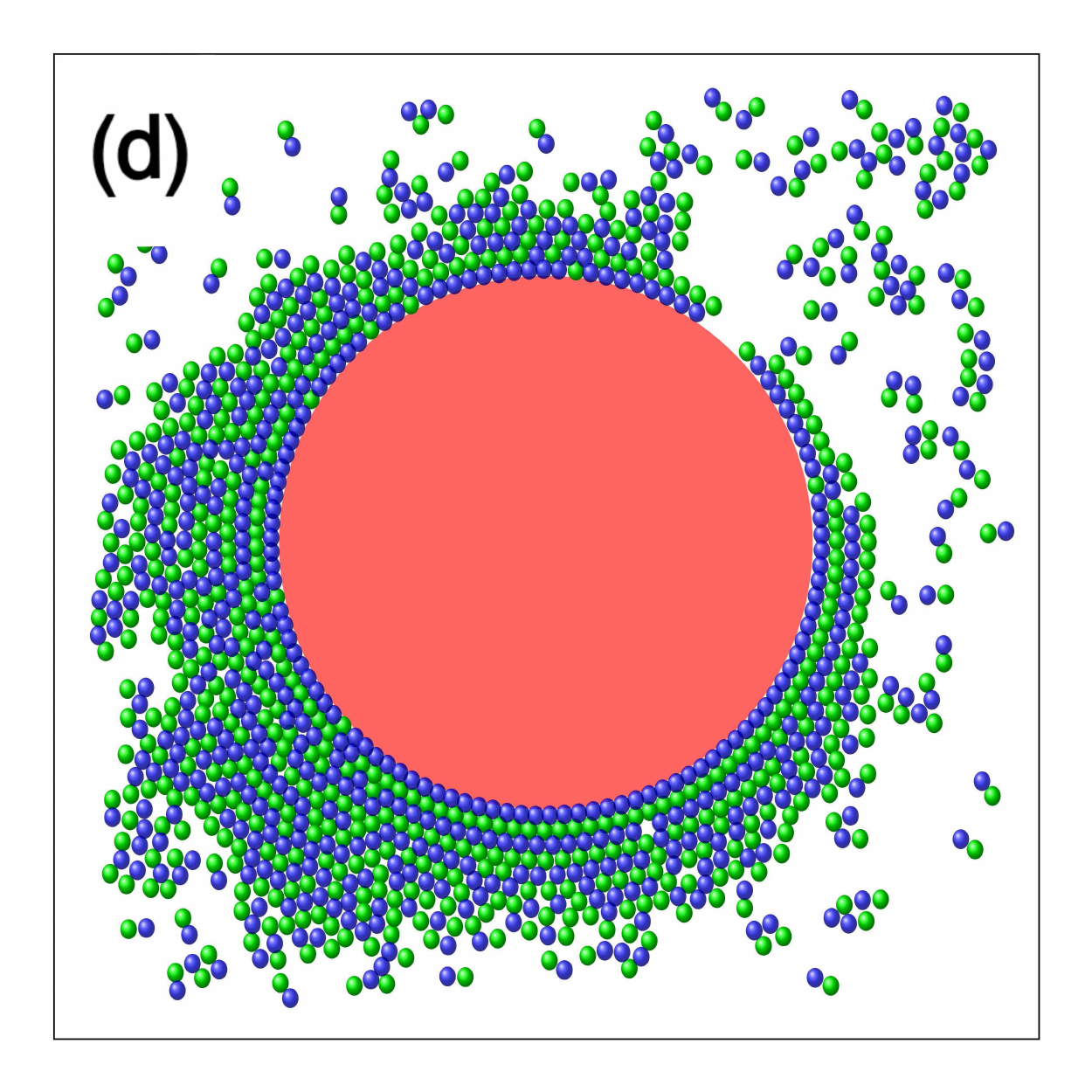}
    \caption{ Simulation Snapshots illustrate aggregation of 
      active dumbbells on the obstacle for various P\'eclet numbers $Pe = 0, 15,20,$ and $40$ (a-d) at   $R_o = 15$. The blue monomer represents the head, the green monomer is the tail of the active dumbbells, and the red disk is the static obstacle.}
    \label{fig:aggregation}
\end{figure}

\begin{figure}[ht]
    \centering
    \includegraphics[width=\columnwidth]{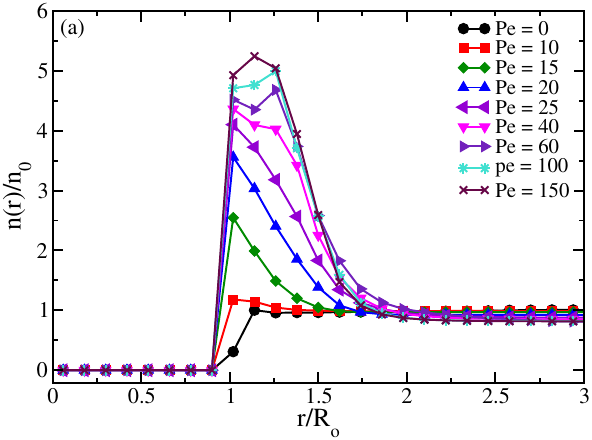}
     \includegraphics[width=\columnwidth]{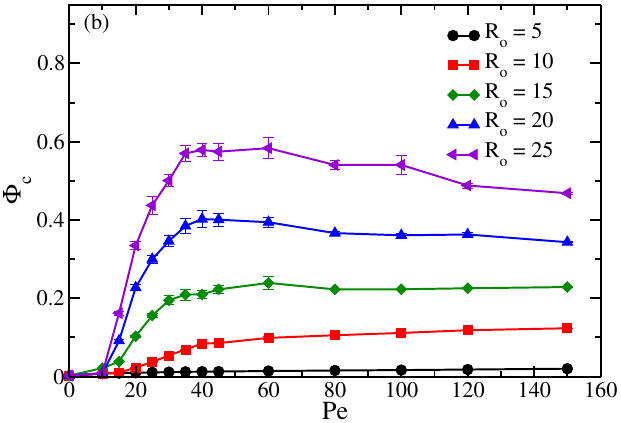}
    \caption{(a) Normalized density distribution  $n(r)/n_0$ of active dumbbells around the center of the obstacle for various $Pe$ at a given $R_{o} = 15$ as a function of $r/R_o$. b) The average fraction of active dumbbells onto the obstacle's surface $\Phi_c = \frac{<N_c>}{N_m}$ with P{\'e}clet number  $Pe$ for various $R_{o}$.}
    \label{fig:cluster_size}
\end{figure}

\begin{figure}[ht]
    \centering
\includegraphics[width=\columnwidth]{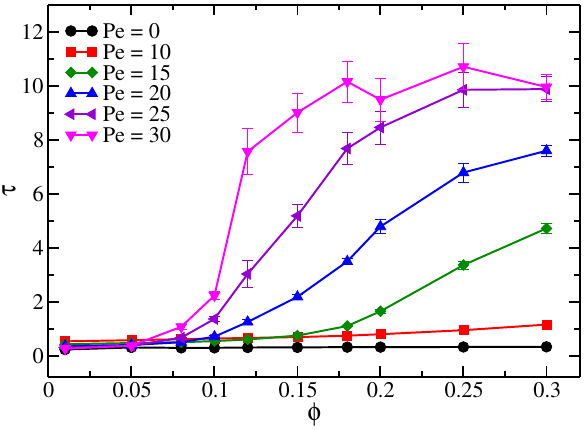}
    \caption{ Plot shows the variation of the residence time $\tau$ of active dumbbell as a function of area fraction $\phi$ for various $Pe$ at $R_o = 15$.}
    \label{fig:retent_time}
\end{figure}

\subsection{Aggregation on obstacle}
 The active dumbbells are uniformly distributed at low P\'eclet numbers with uniform density ($n_0$). However,  as the P\'eclet number increases, active dumbbells accumulate near the interface, leading to aggregation in the form of a large cluster. This aggregation substantially grows with the activity (Pe). A snapshot of active dumbbells with an obstacle illustrates aggregation behavior near the surface for various P\'eclet numbers: $Pe = 0$, 15, 20, and 40,  see Fig.~\ref{fig:aggregation} a-d. The average density profile around the obstacle in terms of the color map is also presented in the SI material in Fig.~\ref{fig:aggregation}.

%{Firstly, we study the structural configuration of the system. In the absence of obstacles, there is also the aggregation of dumbbells that takes place at high activity and above a threshold density\cite{suma2014motility}. In our system, we keep density fixed to $\phi = 0.2$, which is less than the threshold density, so that for all $Pe$ ranges, there is no aggregation without an obstacle. }

To gain insights into the non-uniform density profile near the obstacle, we quantify the density distribution function $n(r)$ as a function of radial distance from the center of the obstacle ($r/R_o$). The spatial variation of the normalized density $n(r)/n_0$ is presented in Fig.~\ref{fig:cluster_size}-a for various $Pe$ at a given $R_o=15$. {\cblue As expected for $r/R_o\leq 1$, the density of the active dumbbells is zero inside the obstacle.} The figure clearly illustrates the spatial variation of $n(r)/n_0$, which was nearly uniform at relatively low P\'eclet numbers $Pe \leq 10$.%; this indicates that active dumbbells are homogeneously distributed in the proximity of the obstacle.
%$n(r)$ is calculated by considering concentric annular rings of width $1.8\sigma$ around the obstacle. 
% For  P\'eclet numbers ($Pe \leq 10$), the density profile is nearly constant beyond the surface of the obstacle). The density becomes larger than the bulk near the surface for larger   P\'eclet numbers ($Pe  >10$).

However, as the P\'eclet number increases ($Pe > 10$),  a peak appears on the normalized density distribution $n(r)/n_0$, which eventually saturates for larger $Pe$. The height and width of this peak exhibit enhancement and subsequent saturation upon increasing the P\'eclet number.  The larger peak height suggests the size of the aggregated cluster grows as depicted in Fig.~\ref{fig:cluster_size} a.  The density profile reaches the bulk value far from the obstacle and remains the same as Fig.~\ref{fig:aggregation}a-d reflects for all $Pe$  beyond $r/R_o >2$.  The slowdown of the active dumbbells on the surface causes the aggregation and vice versa\cite{cates2013active}.  {For $Pe>60$, the peak width starts weakly decreasing with augmentation of $Pe$. }

%Further, the alignment interactions among active dumbbells lead to the nematic ordering in the aggregate.    

%{\cred This behavior arises due to the relatively low activity, allowing dumbbells to move freely and explore the entire space around the obstacle. However, as the P\'eclet number increases ($Pe \geq 15$), the system transitions into an aggregated state. In this state, the density distribution $n(r)$ from the center of the obstacle displays a peak near the obstacle's surface. This peak signifies a higher concentration of dumbbells in close proximity to the obstacle, and the distribution curve decreases as we move farther away from the surface of the obstacle, reflecting a decrease in the density of dumbbells. Additionally, as the P\'eclet number increases, the width of the peak also increases, indicating the growing size of the aggregation, which eventually saturates for high values of $Pe$. }

Now, we compute the fraction of active dumbbells accumulated on the obstacle's surface; for this, only those active dumbbells are considered who are already on the surface or part of some cluster on the surface. An active dumbbell is part of an aggregate if the closest distance between any dumbbell's monomer is less than $1.2$\cite{Spotswood_D_Stoddard}. The computed fraction of active dumbbells  $\Phi_c=<N_c>/N_d$ is plotted in Fig.~\ref{fig:cluster_size}-b as a function of $Pe$ for various $R_o$, where $<N_c>$ corresponds to the average number of active dumbbells aggregated on the obstacle's surface. Interestingly, the aggregation on the obstacle appears beyond a critical radius ($R_o^c \approx 10$),  as Fig.~\ref{fig:cluster_size} illustrates $N_c$ is nearly zero and does not show significant variation with  $Pe$ for $R_o= 5$.  For $R_o\geq10$,  the accumulation of active dumbbells is more prominent in the limit of large $Pe$, as Fig.~\ref{fig:cluster_size}-b depicts. 

Similar behavior has been reported in the experiments, where bacterial cells are entrapped by the convex surface only beyond a critical radius\cite{sipos2015hydrodynamic}. {\cblue For ($R_{o}=10$), the size of aggregation monotonically grows beyond a particular value of $Pe$ before approaching a plateau value in the limit of large $Pe$. 
For larger obstacles $R_o=15$, $20$, and $25$, Fig.\ref{fig:cluster_size}-b displays that the accumulation weakly decreases in the limit of large $Pe$.} %This behavior is due to compact aggregation becoming smaller width; see Fig.\ref{fig:cluster_size}-a.  } 

{\cblue    To show the microscopic origin behind the aggregation beyond a critical radius and $Pe$, we compute the residence time of active dumbbells on the surface as a function of area fraction, incorporating the effect of many-body interactions. The residence time is defined as the average time an active dumbbell spends within a cutoff of $1.5$ unit distance from the surface of the circular obstacle. As expected, the residence time is nearly independent of the speed and curvature in the dilute limit; the computed values for various  $Pe$ and $R_o$ are shown in  SI- Fig.3. Further, the residence time   ($\tau$)  as a function of the area fraction for various $Pe$ is also shown in  Fig.~\ref{fig:retent_time}, which is nearly unchanged in the dilute limit. However, as the area fraction approaches $0.1$, a rapid growth in $\tau$ appears as Fig.~\ref{fig:retent_time} displays. More importantly, $\tau$ has increased nearly two orders of magnitude from its dilute limit. The rapid increase in residence time is attributed to many-body interactions at higher area fractions, indicating the slowdown of rotational diffusion of active dumbbells.  
 
The aggregation mechanism reported here is similar to the MIPS; when an active dumbbell encounters an obstacle, it remains on the surface up to characteristic time $\tau_c \approx 1/D_r$ before reorientation of its axis due to thermal motion. In the meantime, if other active dumbbells collide with this, they get stuck, leading to a gradual decrease in the rotation of its orientation axis, which further attracts more active dumbbells on this aggregate.  This slowdowns the speed of fast-moving dumbbells and their rotational diffusion, instigating positive feedback beyond a critical $Pe$.  This leads to the phase separation of active dumbbells with a high-density phase on the obstacle surface, coexisting with the low-density phase in the bulk.  In summary, we have shown here that aggregation of the active dumbbells on the obstacle occurs only beyond a critical radius and P{\'e}clet number due to the dramatic slowdown of the rotational diffusion.  }

%{\cred To obtain a comprehensive understanding of the aggregate's overall size, we measure the average fraction of dumbbells within the aggregate, defined as $\Phi_c = \frac{<N_c>}{N_d}$, where $N_c$ represents the number of dumbbells that are part of the aggregate, calculated using cluster analysis \cite{Spotswood_D_Stoddard} with a cutoff distance of $1.2\sigma$. The nature of $\Phi_c$ as a function of $Pe$ for different $\sigma_{obs}$ values is illustrated in Fig. \ref{fig:cluster_size} (b). For $\sigma_{obs} \leq 10$, $\Phi_c$ remains nearly zero across all values of $Pe$, indicating the absence of aggregation around the obstacle. This observation is due to the limited surface area available for the aggregation of dumbbells around the obstacle. However, when the surface area becomes sufficiently large ($\sigma_{obs} \geq 20$), we observe a dramatic transition in $\Phi_c$, shifting from zero to a saturated value. This transition corresponds to the shift from a homogeneous to an aggregated state. As $\sigma_{obs}$ increases, the surface area also becomes larger, intensifying the aggregation process, and eventually $\Phi_c$ increases. The aggregation mechanism reported here is quite intuitive: when a dumbbell encounters an obstacle, it remains in close proximity for a characteristic time $\tau_c = 1/D_r$ before reorienting its direction. This slowdown of fast-moving dumbbells due to repulsive interaction with the obstacle triggers a positive feedback loop for aggregation.  } 
\begin{figure}[t]
     \centering     
     \includegraphics[width=\columnwidth]{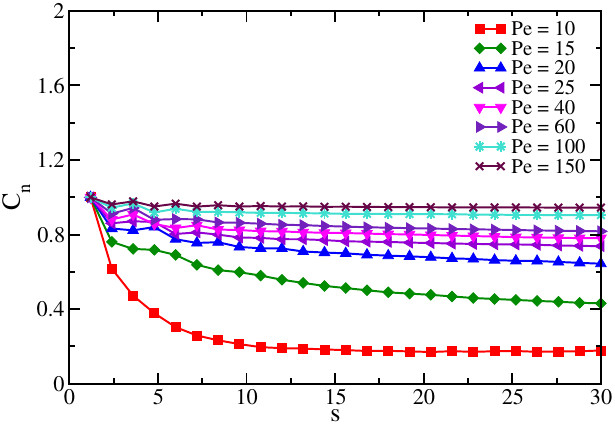}
     \caption{The alignment correlation function $C_n(s)=<\hat{\mathbf{n}}_i.\hat{\mathbf{n}}_j>$ of active dumbbells around the circumference as a function of arc distance (s)   for various $Pe$, at a given $R_o=15$.}
     \label{fig:orient_dist}
 \end{figure}
 
 \begin{figure}[ht]
    \centering
    \includegraphics[width=\columnwidth]{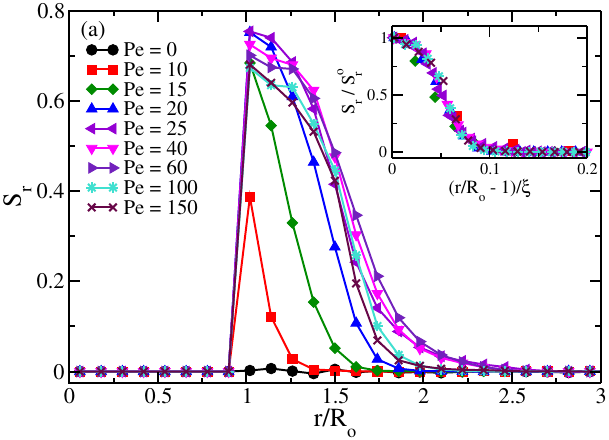}
    \includegraphics[width=\columnwidth]{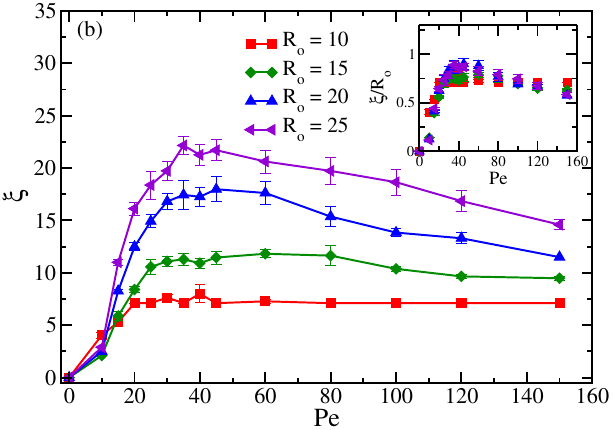}
    \caption{ (a) The local alignment of the active dumbbells w.r.t. normal of the surface ($S_r = <-\hat{\mathbf{n}}.\hat{\mathbf{r}}>$) is plotted as a function of radial distance from the center of the obstacle for various $Pe$ and at $R_o = 15$. {\cblue Inset displays master curve of  local alignment parameter $S_r/S^o_r$ as a function of $(r/R_o-1)/\xi$, where $S^o_r=S_r$ at $r=R_o$ . 
    (b) The variation of the correlation length ($\xi$) is obtained from the radial local alignment parameter $S_r$,  as a function of $Pe$ for various $R_o$. Inset shows the universal correlation length ($\xi/R_o$) with the $Pe$.} }
    \label{fig:cos_correl}
\end{figure}

\subsection{Orientational Ordering}
The excluded volume interaction among anisotropic particles along with the surface causes orientational ordering among active dumbbells in the high-density phase\cite{de1993physics}; therefore, the polar ordering of the active dumbbells in aggregate is inevitable near the surface.  As a result, the ordered phase emerges within the aggregation of active polar dumbbells. 
 Simulation snapshots of Fig. \ref{fig:aggregation} a-d display the tilted alignment and the long-range ordering of active dumbbells along the circumference within a few layers.

 We compute the orientational correlation function, which measures the length scale of ordering along its circumference; for this, we use the following expression, 
\begin{equation}
 C_n (|s-s'|) = <\hat{\mathbf{n}}_i (s) \cdot \hat{\mathbf{n}}_j(s')>,
\end{equation} 
where $s$ and $s'$ are the arc positions of the active dumbbells along the circumference of the obstacle, and $\hat{\mathbf{n}}_i $ represents the unit vector of $i^{th}$ active dumbbell,  pointing from the tail to head monomer.   
The active dumbbells within the first layer, i.e., the cutoff distance 1.5 from the obstacle's surface, are considered to compute this correlation function ($C_n$).  
%the orientational correlation function $C_n(s)$ of active dumbbells as a function of arc distance $s$ along the surface for various P\'eclet numbers. T
Figure \ref{fig:orient_dist} displays alignment correlation $C_n$ decreases rapidly with $s$ for the small P\'eclet numbers $Pe \leq 20$; however, for $Pe > 20$, the correlation does not decay even  $20\%$ of its value for much larger distances. This indicates the presence of the long-range ordering at circumference at higher speeds. 

%The correlation length along the circumference becomes the size of the obstacle radius. 
%and they perform correlated motions, which will be discussed in the latter sections. 

%Importantly, their orientation is not tangential, as seen in the study of the planner wall and in the study of {\it Bacilus  subtillis}\cite{}. 

%{\cred  In order to display the self propulsion ordering near the obstacle, we In the vicinity of the obstacle, arrays of dumbbells align along the curved surface of the obstacle in a tilted manner (see Fig. \ref{fig:aggregation}(d)), resulting in the strongest spatial correlation of dumbbell orientations. Figure \ref{fig:orient_dist} displays the orientational correlation function $C_n(s)$ of active dumbbells as a function of arc distance $s$ along the surface for various P\'eclet numbers.  The cutoff distance for $C_n$ is 1.5$\sigma$ from the obstacle's surface. It is evident that the correlation length (the distance at which the correlation reduces to 1/e) rapidly diverges with increasing $Pe$. } 

 The local alignment of active dumbbells in a radial direction varies with the speed as we go far from the obstacle.  To analyze this, we compute the projection of the orientation vector along the obstacle's radial direction. The radial local alignment parameter is defined as $S_r = -<\hat{\mathbf{n}} \cdot \hat{\mathbf{r}}>$, where  $\hat{\mathbf{r}}$ is the unit radial vector from the center of the obstacle. Figure \ref{fig:cos_correl}-a displays computed values of $S_r$ as a function of radial distance $r$  for various $Pe$ at a given $R_{o} = 15$.  As expected in equilibrium, aggregation is absent; therefore, active dumbbells are randomly orientated, resulting in $S_r$ being nearly zero everywhere. Once the aggregation sets on the surface, for non-zero activity,  $S_r$ increases and exhibits a maximum at the surface.  {\cblue This indicates that alignment becomes prominent for larger $Pe$ at the surface and grows in the radial direction. Strikingly, the height of peak and width weakly decrease for $Pe>60$ as Fig.~\ref{fig:cos_correl}-a shows.   The decrease in the peak and width indicates that the tilt angle increases for $Pe>60$. 
 Further, as one moves away from the surface, $S_r$ rapidly diminishes, implying the random ordering of active dumbbells in outer layers and bulk.  }

 The length scale over which this orientation remains significant from the surface can be computed from the local radial alignment parameter. In equilibrium, near the wall, the ordering function $S_r$ decays exponentially in dense aggregates as $S_r\approx  \exp(-r/\xi)$\cite{Doi2013SoftMP}. However, the exponential decay is limited only to a short range due to the limited aggregate size. Therefore, to estimate the characteristic correlation length ($\xi$), we take the distance over which the orientation curve has reduced to 1/e of its maximum value. {\cblue The estimated correlation length sharply grows with $Pe$  as   Fig.~\ref{fig:cos_correl}-b  depicts. However, beyond a certain $Pe$ for $R_o<15$, the correlation length saturates. The saturation of the $\xi$ is a consequence of the finite size aggregate on the obstacle.  Besides, for $R_o>10$, correlation length displays the non-monotonic behavior where the correlation length decreases for the larger $Pe$, see Fig.~\ref{fig:cos_correl}-b. This observed behavior of the correlation length is consistent with the cluster size and distribution of the active dumbbells around the obstacle, as shown in Fig\ref{fig:cluster_size}-a, and b.  

% This can be further ascertained from Fig.~\ref{fig:cluster_size}-b where $\Phi_c$ has qualitatively similar behavior to $\xi$.
 
 Interestingly, we can find the universal behavior of the correlation length by scaling with the obstacle size $(R_o)$. The inset of Fig.~\ref{fig:cos_correl}-b shows the scaled curve of $\xi/R_o$ with  $Pe$ for various $R_o$. This indicates that the orientational ordering has a length scale of $R_o$. Moreover, using the correlation length,  we can obtain the universal behavior of the local alignment order parameter $S_r$ as presented in the inset of  Fig.\ref{fig:cos_correl}-a. This displays the $S_r/S_r^o$ also attains a master curve if presented as a function of $(r/R_o-1)/\xi$ for various $R_o$ where $S_r^o$ is the value of the $S_r$ at $r=R_o$. }

%This high degree of orientation near the obstacle's surface, with $S_n$ rapidly diminishing as one moves away from the obstacle's surface. Furthermore, with the increase of $Pe$, the width of the $S_n$ curve also increases, indicating the increasing length scale of orientation from the surface. To ascertain the distance over which this orientation remains significant, we calculate the characteristic length, defined as the distance at which the orientation curve decays to its 1/e value, as a function of $Pe$ for various values of $\sigma_{obs}$, as illustrated in Figure \ref{fig:cos_correlation}. At a fixed $\sigma_{obs}$, the characteristic length initially increases with rising $Pe$ but later saturates due to the limitation imposed by the size of the aggregation. The characteristic length also increases with an increase in $\sigma_{obs}$ because of the larger aggregation size resulting from a greater surface area available at higher $\sigma_{obs}$.

\begin{figure}[t]
    \centering
    \includegraphics[width=0.98\columnwidth]{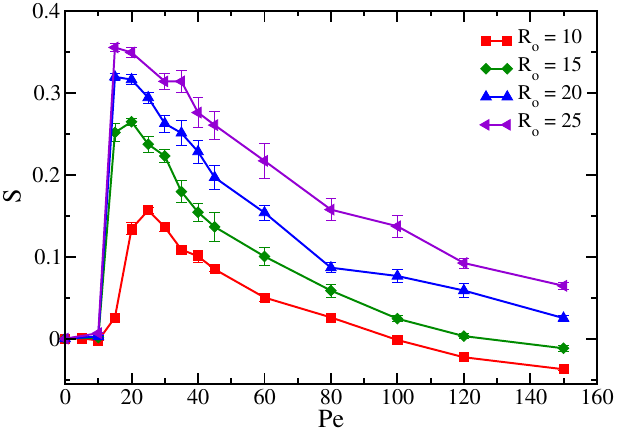}
    \caption{Average orientational order parameter $S = <2\cos^2\theta - 1>$  as a function of $Pe$ for various $R_{o}.$  Here $\theta$ is the orientation of active dumbbells from the normal to the obstacle's surface.}
    \label{fig:order_para_pe}
\end{figure}

The average global orientational order parameter quantifies the ordering within the system; this can be defined as $S = <2 ({\bf \hat n }\cdot {\bf \hat r})^2 - 1> =  <2 \cos^2(\theta) - 1> $, where $\theta$ represents the orientation of active dumbbell from the normal to obstacle's surface. Figure~\ref{fig:order_para_pe} displays the computed values of global orientational order parameter $S$, which increases sharply from zero to a non-zero value beyond a critical value of $Pe$. This suggests spontaneous ordering and aggregation of active dumbbells on the surface. After approaching a maximum,  $S$  starts to descend with $Pe$. The decrease in global orientational order parameter could be understood from the dynamic behavior of the active dumbbells near the obstacle. The active dumbbells form a tilted array around the circumference of the obstacle; the size of the array increases with $Pe$. The rotational motion of the aggregation also strengthens with $Pe$, resulting in a larger tilt angle $\theta$ and, consequently, a reduction in the global order parameter ($S$). {\cblue The larger tilt angle drives active dumbbells to slide away from the cluster in the limit of large $Pe$, hence resulting in the decrease of the cluster size and its width for larger activity as has been seen in Fig.\ref{fig:cluster_size}-(a) and (b). This also causes non-monotonic behavior of the correlation length $\xi$ in the same limit. } 
%Moreover, the increase in $S$  with  $R_o$ outcomes from larger aggregation on the bigger obstacle.

In summary, domains with locally ordered dense phases are prominent near the surface and coexist with the randomly orientated dilute phase far from the surface. The ordered phase on the surface appears beyond a critical radius $R_c$ and $Pe$. The dynamical characteristics of the active dumbbells are provided in the subsequent section. 

 %Furthermore, as discussed in the previous section, active dumbbells align consistently within the aggregate near the obstacle's surface, and as we move away from the surface, the orientations of the dumbbells become a random distribution.   Therefore, the alignment interactions primarily dominate the inner portion of the aggregation about the obstacle's circumference. 

\begin{figure}[h!]
    \centering
      \includegraphics[width =\columnwidth]{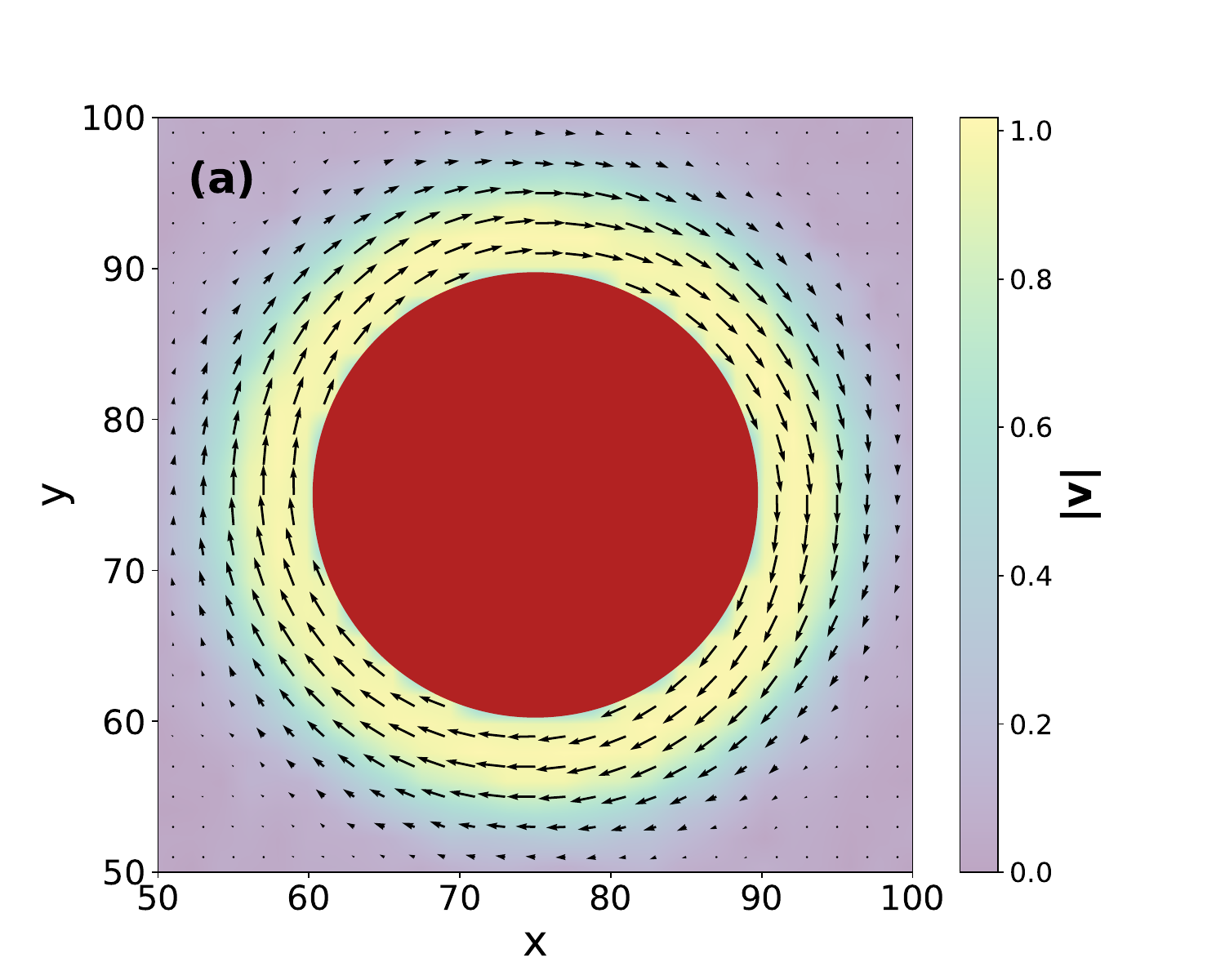}
      \includegraphics[width=\columnwidth]{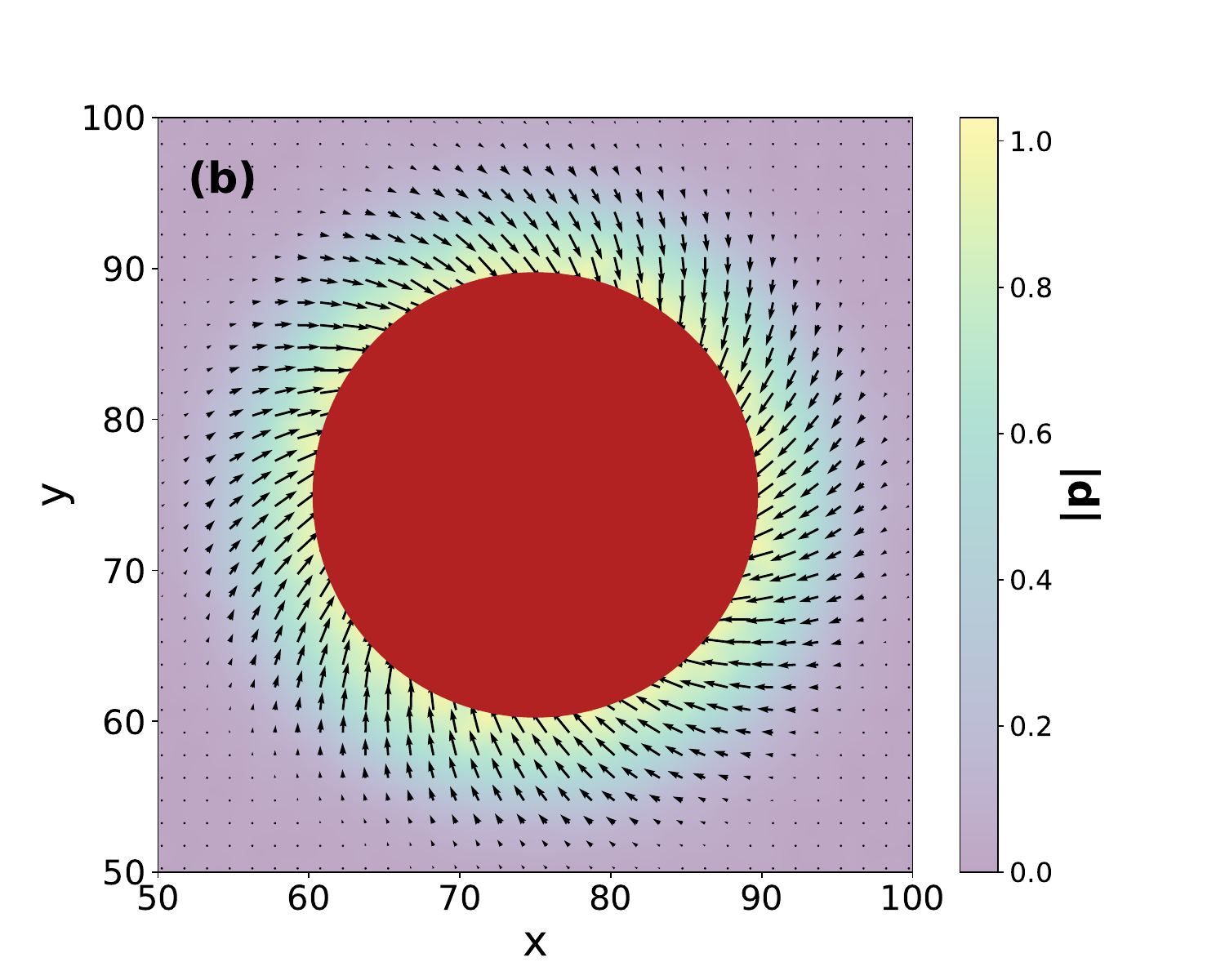}
    \caption{(a) The normalized local velocity field $\mathbf{v}$ and (b) normalized polarization  vector field ${\mathbf{p}}$ of active dumbbells at a given obstacle radius  $R_o = 15$ and activity $Pe = 40$. The color bar indicates the magnitude of the $\mathrm{v}$ and $\mathrm{p}$ in the respective plots. }
    \label{fig:velo_map}
\end{figure}

\begin{figure}[ht]
    \centering
   \includegraphics[width=\columnwidth]{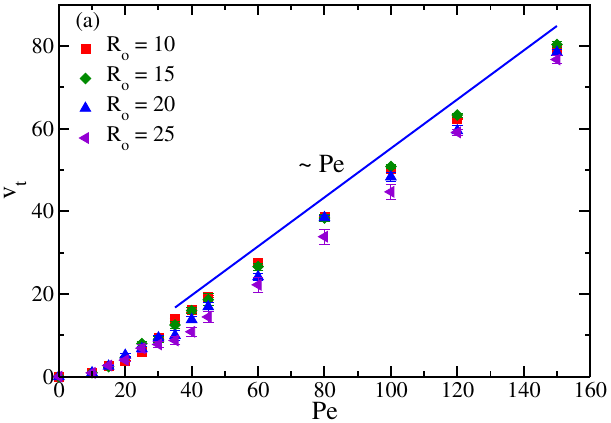}
        \includegraphics[width=\columnwidth]{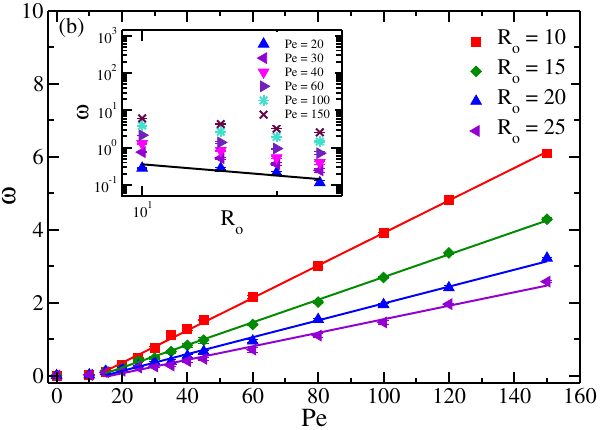}
    \caption{ (a) Variation of the average tangential speed $\mathrm{v}_t$ of the active dumbbell as a function of $Pe$ for various $R_o$. (b) The variation of average angular speed $\omega$ of the active dumbbell as a function of $Pe$ for various $R_o$. This shows tangential and angular speeds linearly grow with $Pe$ in the dense phase. Solid lines show the linear behavior of the angular and tangential speeds.  The inset shows the variation of rotational speed $\omega$ with $R_o$ for different $Pe$. {\cblue The black solid line shows the power behavior of the $\omega$ given as $\omega \sim R_o^{-1}$.}}
    \label{fig:tan_vel_dum}
\end{figure}

\begin{figure}[ht]
    \centering
    \includegraphics[width=\columnwidth]{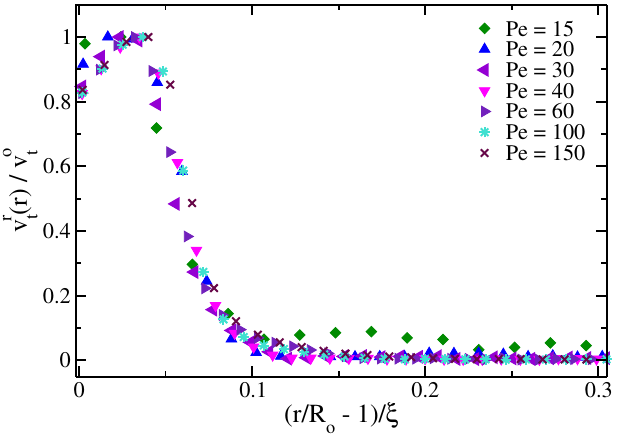}    
    \caption{ {\cblue The scaled tangential speed $\mathrm{v}_t^r(r) / \mathrm{v}_t^o$ of active dumbbells as a function of scaled linear distance $(r/R_o - 1)/\xi$ from the centre obstacle for various $Pe$ at a given radius $R_o$ =15.}}
    \label{fig:tan_vel_dis}
\end{figure}

\subsection{Dynamics of active dumbbell}
The aggregate of active dumbbells spontaneously rotates around the obstacle beyond a critical radius and P\'eclet number. The dynamic behavior of the active dumbbells is captured in various supporting movies; see SI -Movie 1, 2, and 3 for various P\'eclet numbers at $R_o=15$ and Movie 3, 4, and 5 illustrates for various radii at a fixed $Pe=40$.  The persistent rotational motion of the active dumbbells occurs due to orientational ordering and tilted alignment. The rotational motion can also be visualized from the average streamline plot displayed in Fig.~\ref{fig:velo_map}-a, which illustrates the circulatory flow profile of the velocity vector around the obstacle.     
%leading to the tangential force that applies the torque on the aggregate. The resultant torque drives aggregate to rotational motion, forming a vortex. 

Notably, to demonstrate the direction of the velocity profile $\mathbf{v}$ and local polarity direction  ${\mathbf{p}}$ are not necessarily the same, we present local polarization and velocity vectors in  Fig.~\ref{fig:velo_map} as a vector map. The average vector map  ${\mathbf{p}}$ illustrates the spiral form on the surface, where they all converge on the obstacle at some tilted angle.  The vorticity of the velocity field $\mathbf{v}$ and spiral form of the local polarisation vector ${\mathbf{p}}$ on the surface ensures the rotational motion of the aggregate.   

To quantify the rotational motion of the cluster, we compute the tangential component of the velocity of active dumbbells. The tangential speed of active dumbbell w.r.t. the obstacle is ${\mathrm{v}}_t = <{\bf v}\cdot \hat \theta >$, $\hat \theta$ is the unit vector in the azimuthal direction from the center of the obstacle. 

The variation of the average tangential speed, $\mathrm{v}_t$, of a dumbbell within the aggregate with P\'eclet number ($Pe$) is depicted in Fig.~\ref{fig:tan_vel_dum}-a. The tangential speed $\mathrm{v}_t$, for a given $R_{o}$ beyond $Pe>10$,  grows  linearly with $Pe$. Moreover, it shows weak dependence on the obstacle radius $R_o$ in the limit of small $Pe$.  The slope of $\mathrm{v}_t$ weakly decreases with $R_o$, indicating a decrease of tangential speed ($\mathrm{v}_t$) in the limit of large $Pe$ for larger radius ($R_o$).  The lower tangential speed for a more considerable radius $R_o$ is due to the large aggregates on the obstacle's surface, which causes a slowdown.  This can also be seen in  SI Fig. 2-a, where the average directed speed of active dumbbell decreases for larger aggregates. The average angular speed of the active dumbbell can be estimated from the following expression,
\begin{equation}
   \omega = \left< \frac{1}{|\mathbf{r}|} \left[\hat{\mathbf{r}} \times \mathbf{v}\right] \right>
\label{Eq:omega}.
\end{equation}

As expected, the estimated angular speed $\omega$ also displays a linear increase with $Pe$, see Fig.\ref{fig:tan_vel_dum}-b.  Note that below $Pe<15$, the spontaneous rotation of the cluster does not appear; thereby, the tangential speed and rotational frequency are nearly zero.
Further, we show the dependence of the angular speed $\omega$ on the obstacle radius $R_o$ in the inset of Fig. \ref{fig:tan_vel_dum}-b.   The angular speed exhibits power-law variation given as $\omega \sim R_o^{-1}$. 

%The decrease of the angular speed with the cluster's radius has been reported in previous studies of the active dumbbells in the bulk\cite{suma2014motility} where the angular velocity of the rotating cluster decreases with the size of cluster given as  $\omega \sim R^{-1}$, where $R$ is the radius of the cluster. On the contrary, in our study, the exponent $\beta$ lies in the range of ${1}$ to $1.3$; this change may arise due to larger aggregation fluctuations at higher P{\'e}clet numbers $Pe$. 

{\cblue The local tangential speed of the active dumbbell can be presented as a function of the radial distance.  Assuming the linear behavior of the angular and tangential speeds, we present the local tangential speed as $\mathrm{v}_t^r(r) / \mathrm{v}_t^o$ as a function of  $(r/R_o - 1)/\xi$,  where $\xi$ is correlation length.  Remarkably, the scaled quantity $\mathrm{v}_t^r(r) / \mathrm{v}_t^{o}$ exhibits universal behavior  as depicted in Fig.~\ref{fig:tan_vel_dis}. Here, $\mathrm{v}_t^o$ represents the maximum value of $\mathrm{v}_t^r(r)$ and exhibits a linear dependence on $Pe$.
Surprisingly,  the tangential speed ($\mathrm{v}_t^r(r)$)  displays a non-monotonic variation,  where an increase at shorter distances is followed by a sharp decrease far from the surface, see Fig.~\ref{fig:tan_vel_dis}. The lower tangential speed of the innermost layer results from the dynamic layer coming in contact with the static surface, therefore imposing a higher viscous drag on the inner layers. However, moving away from the surface increases due to a strong alignment correlation in the inner layers. Further away from the surface, The alignment interaction weakens in the cluster; hence, the tangential speed also diminishes in the outer layers.}%  The non-zero value of tangential speed provides torque to the clusters that lead to the rotational motion around the obstacle. } 
%On the other hand, the motion of active dumbbells and alignment is random on the outer layers of the aggregate; therefore, their speed is random, and hence, the tangential component diminishes. 

%This can be further seen in Fig.~\ref{fig:tan_vel_dis} that below the critical P\'eclet number, the tangential speed   $\mathrm{v}_t\approx 0$. 

In summary, we have shown that the front monomer of the dumbbell points toward the obstacle's surface in the aggregate at a certain angle.  The locked orientation of the active dumbbells in the cluster applies local force, which translates into torque and causes the cluster to rotate. The rotation of the cluster can be either anticlockwise or clockwise, depending on the tilt angle of the active dumbbells on the surface. For planar confinement, a similar effect has been reported for the sliding motion of hedgehog clusters of self-propelled rods\cite{wensink_and_lowen_2008}.

\begin{figure}[t]
    \centering
    \includegraphics[width=\columnwidth]{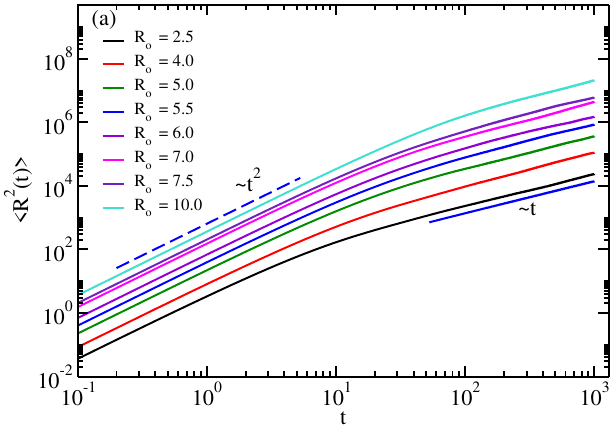}
    \includegraphics[width=\columnwidth]{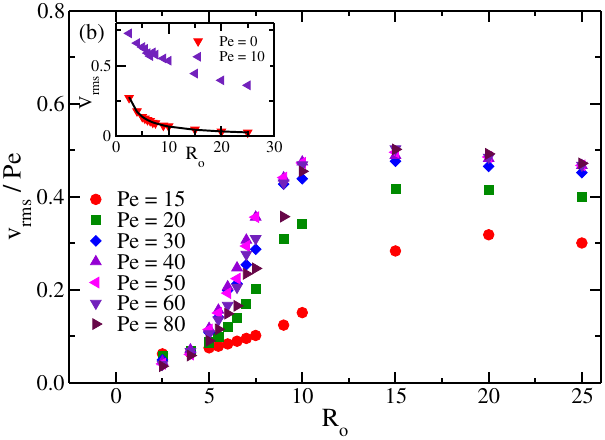}
    \caption{ (a) {\cblue The plot shows  MSD ($<R^2(t)>$) of the tracer particle for different radii  $R_o$ at a given $Pe = 40$, where blue dashed line indicates ballistic behavior $t^2$ and solid line shows the diffusive behavior $t$ of the MSD.} (b) {\cblue The normalized average active speed $\mathrm{v}_{rms}/Pe$ of the circular tracer particle is presented as a function of its radius  $R_o$ for various $Pe$. The inset shows the average speed at $Pe=0$ and $Pe=10$ as a function of $R_o$. For $Pe = 0$, average speed obeys $\mathrm{v}_{rms} \sim R_o^{-1}$ relation shown by solid black line.} }
    \label{fig:msd}
\end{figure}

\subsection{Dynamics of Circular Passive Tracer in Active Medium}
\begin{figure}[t]
    \centering
    \includegraphics[width=\columnwidth]{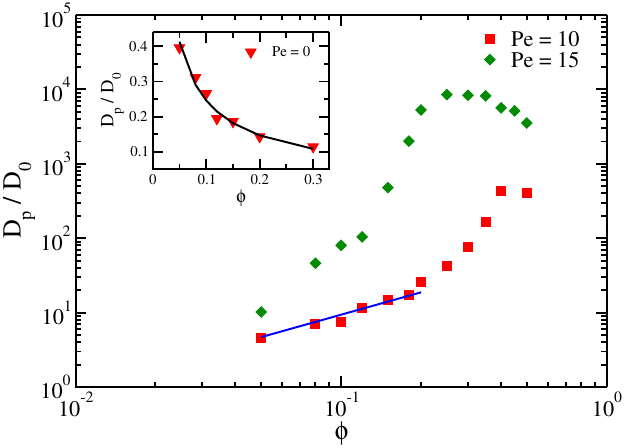}
    \caption{{\cblue The effective diffusivity $D_p/D_0$ of the tracer particle is shown as a function of area fraction $\phi$ for different $Pe$ at a fixed $R_o = 15$, where the diffusion coefficient of the passive tracer is given as, $D_0=k_BT/\gamma_o$. The blue line shows a linear fit for lower values of $\phi$. 
    The inset shows the effective diffusivity of a tracer particle $D_p/D_0$ decreases monotonically with the increase in $\phi$ for the passive bath ($Pe = 0$). }}
    \label{fig:eff_diff}
\end{figure}
We now consider a case where a passive tracer moves in response to thermal and active stresses arising from the medium.  %Due to the asymmetric aggregation of dumbbells, illustrated in Fig. \ref{fig:aggregation} (b)-(d), the net force experienced by the obstacle due to the self-propulsion of dumbbells is non-zero. 
The equation of the motion of the  passive tracer is governed  by the under-damped Langevin equation,
\begin{equation}
    M_o\frac{d\mathbf{v_o}}{dt} = -\gamma_o\mathbf{v_o} + \mathbf{F}_w + \mathbf{F}^R.
\label{Eq:tracer}
\end{equation}
Where $M_o$ represents the mass of the tracer particle, $\mathbf{v_o}$ is the velocity, $\gamma_o$ is the drag, $\mathbf{F}_w$ is the interaction force with the active dumbbells, and $\mathbf{F}^R$ is the random force of a zero mean and correlation given as $<\mathbf{F}^R(t)\cdot {\mathbf F}^R(t')> = 4 k_BT \gamma_o \delta(t-t')$. The mass of the tracer particle is scaled in units of the mass of the dumbbell's monomer $m$ while keeping mass density fixed, with $ M_o = \sigma_o^2 m/(\sigma^2)$.  This leads to the choice of the drag coefficient of the tracer being $ \gamma_o = \gamma_t \sigma_{o}/\sigma$, where $\sigma_o$  is the diameter of the tracer particle. We solve Eq.~\ref{Eq:tracer} using the velocity-Verlet algorithm\cite{allen2017computer}, with time steps ranging from $10^{-4}\tau$ to $5\times10^{-6}\tau$. Other parameters remain same, as listed in the simulation model.

{\cblue To characterize the dynamics of the passive tracer particle in the active medium, we compute the mean-square-displacement (MSD),
\begin{equation}
<R^2(t)> =  <(\mathbf{r}(t ) -\mathbf{r}(0))^2> .  
\end{equation}
{\cblue  As displayed in Fig.~\ref{fig:msd}-a, The MSD of the passive tracer in the active medium attains the superdiffusive behavior, indicated by $<R^2(t)> \sim  <v_o^2> t^2$, which is followed by the diffusive behavior in the long time limit
($<R^2(t)>  = 4 D_p t$) for various values of $R_o$,  at  $Pe = 40$. Where $D_p$ is the effective diffusion coefficient of the tracer particle of the radius $R_o$.}

From the MSD curves, we estimate the average speed of the tracer particle in the superdiffusive regime using the following expression $<R^2(t)> = <v_o^2>t^2$. The average speed of the tracer particle can be computed $\mathrm{v}_{rms} = \sqrt{<v_o^2>}$, which is displayed in Fig.~\ref{fig:msd}-b for various $Pe$. 

For the passive system, i.e., $Pe = 0$, the speed of the tracer particle decreases with $R_o$ following the relation $\mathrm{v}_{rms} \sim R_o^{-1}$. This is intuitive as we know the speed of a tracer $\mathrm{v}_{rms}\sim M_o^{-1/2} $, that leads to $\mathrm{v}_{rms}\sim R_o^{-1} $. For $Pe=0$, a solid line in the inset of Fig.~\ref{fig:msd}-b establishes the speed decrease with the inverse of the obstacle radius.  

{\cblue There is a noticeable shift in the behavior of $\mathrm{v}_{rms}$ with $Pe$ in the active medium; surprisingly,  the speed of tracer particle increases with  $R_o$ which is a very different in passive medium.  More-importantly,  for the $Pe\geq30$, $\mathrm{v}_{rms}$ speed increases with $R_o$ latter it decrecreas at higher packing, see Fig.\ref{fig:msd}-b. The increase in the speed of the tracer particle is a consequence of the increase in the heterogeneous aggregation of the active dumbbells on the surface of the tracer particle with the size of the obstacle, see Fig.\ref{fig:aggregation} and \ref{fig:cluster_size}. This can also be seen in the SI- movies 6 and 7 for P\'eclet numbers $Pe = 10$ and $40$ at fixed $ R_o = 15$. For $R_o>15$, as expected,  the speed of the tracer decreases due to larger viscous drag. Interestingly, beyond $Pe\geq30$, all the curves of the $\mathrm{v}_{rms}$ superimpose on top of each other once they are scaled with the P\'eclet number  (Pe) as Fig.\ref{fig:msd}-b displays the universal behavior of $\mathrm{v}_{rms}/Pe$ with $R_o$.

In the large time regime, the MSD of the tracer particle assumes diffusive behavior as displayed in Fig.~\ref{fig:msd}-a. We estimate the effective diffusion coefficient by varying the area fraction of the active dumbbells using the relation $<R^2(t)>=4D_p t$. The computed values of the effective diffusivity $D_p (\phi)$ as a function of area fraction $\phi$ are depicted in Fig.~\ref{fig:eff_diff}. As expected in the passive bath, the diffusion coefficient monotonically decreases with $\phi$, see in the inset of Fig.\ref{fig:eff_diff}. Interestingly, The effective diffusion coefficient of the tracer particle in the active bath with $\phi$ linearly increases in the dilute regime, followed by a rapid upsurge up to $\phi<0.3$, unlike the passive medium.
 Notably, it has increased by over four orders of magnitude. Further, an increase in $\phi>0.3$ causes a decrease in $D_p$. Thus, effective diffusivity also exhibits a non-monotonic behavior with $\phi$; as Fig.~\ref{fig:eff_diff} illustrates. The linear growth of $D_p$ with $\phi$ in the dilute regime and non-monotonic feature is consistent with experimental observations\cite{mino2011enhanced,jepson2013enhanced,leptos2009dynamics,patteson2016particle}.}
 In summary,  the increase in the effective diffusion results from active noise in the dilute and intermediate packing fractions. The decrease of the effective diffusivity at higher density is due to the slowdown of the speed of the active dumbbell and the formation of larger aggregation in the regime.  }

\section{Conclusion}
This article presented a comprehensive study of active dumbbells in the presence of a circular static obstacle. We have shown that beyond a critical obstacle radius $R^c_o \approx 10$ and P\'eclet number, active dumbbells aggregate on the surface of the static obstacle. This aggregate spontaneously rotates along the obstacle circumference. The average tangential speed and rotational frequency of active dumbbells within the aggregation increase linearly with the P\'eclet number. The rotation of the aggregate emerged from the torque imposed by the jammed orientation of active dumbbells in the cluster. We have also shown that the local tangential speed of the active dumbbells reflects the non-monotonic behavior as one moves away from the surface. More importantly,  all the curves show universal behavior. The tangential speed initially increases but sharply decreases in the outer layers. The former is a consequence of the jamming of active dumbbells on the innermost layer of aggregate, while in the outer layers, active dumbbells fluctuate; therefore, tangential speed decreases. 

To understand the microscopic origin of the surface aggregation, we also analyzed the residence time $(\tau)$ of active dumbbells in the proximity of the obstacle as a function of area fraction and its speed. A rapid increase in the residence time is shown with a slight increase in concentration. This non-linear increase in the residence time is attributed to the slowdown of rotational diffusion, which leads to the aggregation and, thereby, alignment of the active dumbbells on the surface. This provides the microscopic origin of the aggregation and rotational motion and its qualitative dependence on the curvature radius and the activity.

In addition, we have shown that the average polarization vector $ \mathbf{p}$ differs from the velocity vector profile. The former has a spiral-like profile, while the latter displays a vortex form. The observed distinct directions of the local polarization vector with the velocity field have also been reported in studies of active dumbbells in the absence of solid interface\cite{suma2014motility}; this mandates the direction of the velocity profile, and the polarization vector needs to be treated separately in the microscopic theories\cite{de1993physics,suma2014dynamics}.  Furthermore, our study also reveals a strong orientational ordering among active dumbbells within the aggregate and along the circumference of the obstacle. The active dumbbells align themselves on the surface at a certain orientational angle from the normal to the surface; this angle further decreases with obstacle radius and increases with $Pe$. Moreover, the correlation length of the orientational alignment demonstrates a long-range ordering, with the correlation length becoming as large as the size of the aggregate.  
{\cblue Additionally, we have shown the universal behavior of local alignment  ($S_r$) and local tangential speed  ($\mathrm{v}^r_t$) in terms of the correlation length ($\xi$) that itself grows linearly with $R_o$.

 Our simulations demonstrate a particular case when obstacles are free to move in response to active and thermal forces. The tracer particle displays directed motion. Moreover, the tracer particle's speed increases with its size, followed by a decrease in the limit of the larger radius beyond a critical P\'eclet number, despite the increase in the viscous drag with size. Furthermore, we have also shown that the effective diffusivity of the tracer particle displays the non-monotonic behavior with the density of the active dumbbells in the active medium.  This non-intuitive increase in the speed and diffusion coefficient is a consequence of a large aggregation of active dumbbells on the surface of the tracer particle compared to smaller ones. The non-monotonic behavior of the effective diffusion of the tracer is also observed in the bacterial solution\cite{patteson2016particle}. This indicates the transport of the relatively bigger particles in the active medium is relatively easier, which can have potential applications in the targeted delivery of the drugs and extracting work from the active medium\cite{nishiguchi2018engineering,das2020aggregate}.  }

 Previous studies have revealed that non-uniform aggregation of active particles on the surface of the asymmetric tracer that guides to directed \cite{bechinger2016active,mallory2014curvature,kaiser2013capturing,smallenburg2015swim} as well as rotational motion\cite{angelani2009self,di2010bacterial,sokolov2010swimming}.
 Recent work has shown that such motion is not restricted to asymmetric tracer particles; a symmetric tracer with critical porosity in the active solution also exhibits directed motion\cite{das2020aggregate}. On the contrary, our study extends this behavior to a smooth, symmetric system. It unveils that the directed motion can be achieved for the relatively bigger-sized symmetric tracer particle in the solution of active dumbbells. 
 Our results will be helpful in elucidating the behavior of the microbial dynamics in microfabricated devices\cite{nishiguchi2018engineering,wioland2016ferromagnetic,pietzonka2019autonomous}.

% Our study demonstrates that directed motion can be obtained using larger symmetric tracer particles in an anisotropic active dumbbell solution.

 The surface-driven aggregation and vortex formation have promising applications in controlling the turbulent dynamics of active suspensions\cite{nishiguchi2018engineering,wioland2016ferromagnetic,dombrowski2004self,genkin2017topological,zhou2017lyotropic,qi2022emergence}.  The hydrodynamic interactions are ignored for the model's simplicity, albeit they may have a considerable impact on the entrapment of the microswimmers\cite{mokhtari2017collective,pan2020vortex,wang2022obstacle}.  However, their influence is not expected to alter the qualitative behavior of the collective dynamics of the active dumbbells described in this study. {\cblue Our model considers a simple model that ignores the influence of deformation of the microswimmer in the dense medium. This may play a critical role in the aggregation and entrapment near interface\cite{mokhtari2017collective,palacci2015artificial,wang2022obstacle}.
    To closely align with the experimental investigations of bacterial dynamics, a periodic array of obstacles, incorporating the hydrodynamic interactions, may be considered in future works\cite{nishiguchi2018engineering,wioland2013confinement,chopra2022geometric}.   }

\section{Acknowledgments}
SPS  and CHT acknowledge financial support from the DST-SERB Grant No. CRG/2020/000661 and UGC, respectively. The computational facilities of IISER Bhopal and Paramshivay NSM  at IIT-BHU are highly acknowledged.
%The considered model has been extensively studied for active spherical systems in the presence of static obstacles, but there is a significant gap in understanding how anisotropic-shaped active particles behave in such scenarios. 

% {\cblue }

\balance

%If notes are included in your references you can change the title from 'References' to 'Notes and references' using the following command:
%\renewcommand\refname{Notes and references}

%%%REFERENCES%%%
%\bibliography{reference} %You need to replace "rsc" on this line with the name of your .bib file
\bibliographystyle{rsc} %the RSC's .bst file
\providecommand*{\mcitethebibliography}{\thebibliography}
\csname @ifundefined\endcsname{endmcitethebibliography}
{\let\endmcitethebibliography\endthebibliography}{}

 \end{document}

%\preprint{AIP/123-QED}

\title[]{Supplementary Information: Collective dynamics of active dumbbells near circular obstacle}% Force line breaks with \\
%\thanks{Footnote to the title of the article.}

\author{Chandranshu Tiwari}
\email{chandranshu21@iiserb.ac.in}
% \affiliation[]{Department of Physics,\\ Indian Institute of Science Education and Research, \\Bhopal 462 066, Madhya Pradesh, India}%Lines break automatically or can be forced with \\
\author{Sunil P. Singh}%
 \email{spsingh@iiserb.ac.in}
\affiliation{ Department of Physics,\\ Indian Institute of Science Education and Research, Bhopal 462 066, Madhya Pradesh, India}%

%\author{C. Author}
% \homepage{http://www.Second.institution.edu/~Charlie.Author.}
%\affiliation{Second institution and/or address%\\This line break forced% with}%

%\date{\today}% It is always \today, today,  but any date may be explicitly specified

 \maketitle

%\begin{quotation}
%The ``lead paragraph'' is encapsulated with the \LaTeX\ \verb+quotation+ environment and is formatted as a single paragraph before the first section heading. (The \verb+quotation+ environment reverts to its usual meaning after the first sectioning command.) Note that numbered references are allowed in the lead paragraph. The lead paragraph will only be found in an article being prepared for the journal \textit{Chaos}.
%\end{quotation}

\section{Normalised Density Map}
\begin{figure}
    \centering
    \includegraphics[width=\columnwidth]{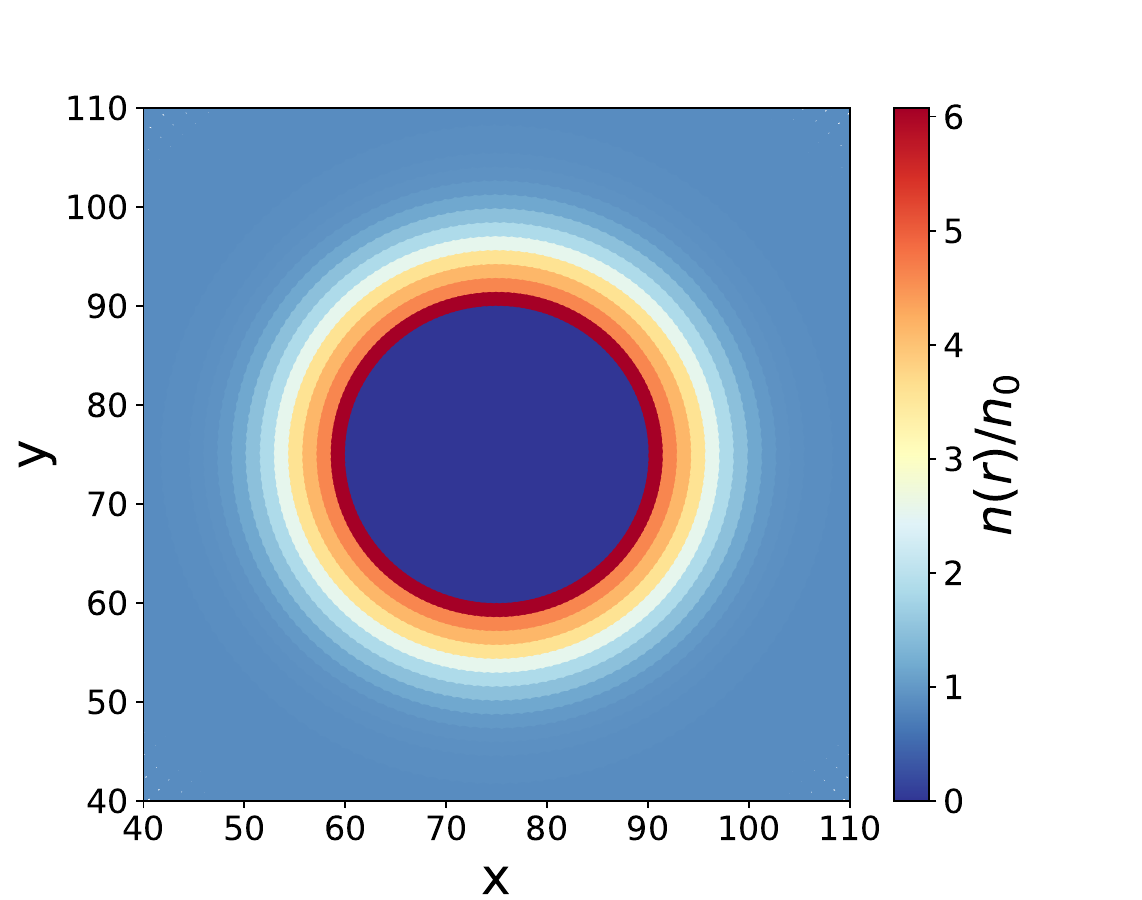}
    \caption{The color map reflects the average density of active dumbbells around the obstacle of radius  $R_o = 15$  at a given $Pe = 40$. The central dark blue patch depicts the circular obstacle with zero density of the active dumbbells. The color bar shows the variation of the normalized density around the obstacle's surface. Only a portion of the simulation box is displayed to clarify the density profile. }
    \label{fig:density_map}
\end{figure}

The normalized average density profile of active dumbbells, represented by $n(r)/n_0$  shown in Figure \ref{fig:density_map}, exhibits a distinctive pattern. Near the obstacle's surface, $n(r)/n_0$ has a maximum value, signifying the highest concentration of active dumbbells. As one moves away from the surface, $n(r)/n_0$ gradually decreases, and far from the surface, the density of the active dumbbells approaches unity, i.e., the bulk value. This indicates that near the surface, the density profile is significantly higher than the bulk, displaying an aggregation of active dumbbells on the obstacle's surface.

\begin{figure}
    \centering
     \includegraphics[width=\columnwidth]{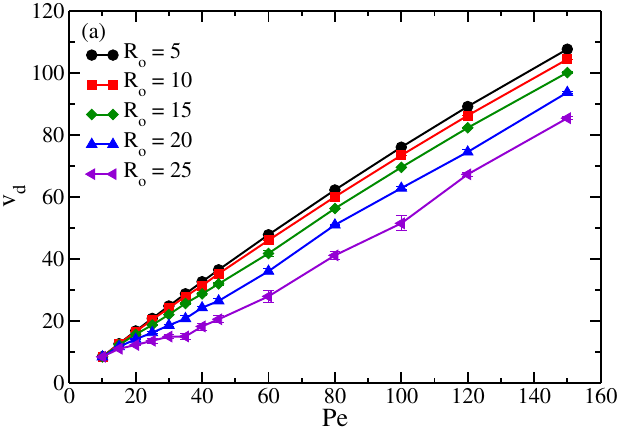}
    \includegraphics[width=\columnwidth]{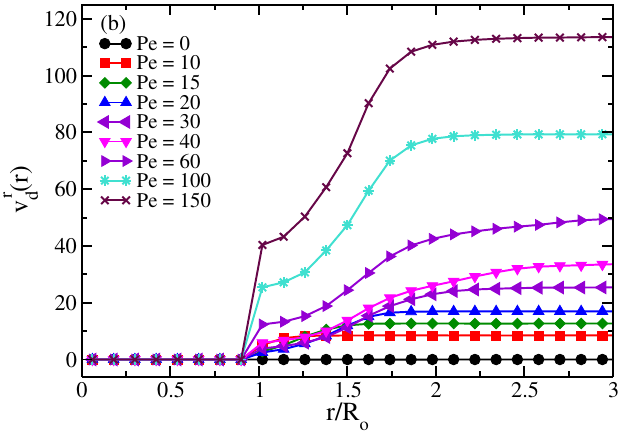}
    \caption{(a) The variation of average directed speed,  as a function of $Pe$ for different radii ($R_o$). (b) The distribution of average directed speed $\mathrm{v}_d^r(r)$ from the center of obstacle for various $Pe$ at a given $R_{o}$ = 15.}
    \label{fig:dir_speed_pe}
\end{figure}

\section{Directed Speed}
 To provide detailed insights into the dynamics of a single active dumbbell, we analyze the average directed speed of a dumbbell in the presence of a static obstacle. The average directed speed ($\mathrm{v_d}$) is defined as the average projection of the velocities of active dumbbells along their axes of orientations, i.e., $\mathrm{v_d} = < \mathbf{v} \cdot \hat{\mathbf{n}}>$. Figure ~\ref{fig:dir_speed_pe}-a displays the average directed speed of a dumbbell as a function of P\'eclet number ($Pe$) for various obstacle radii. Directed speed also grows linearly with $Pe$ at a fixed obstacle radius $R_o$; however, as $R_o$ increases, the $\mathrm{v_d} - Pe$ curve shifts downwards, reflecting the decrease in the directed speed. This is because aggregation size becomes more prominent for higher $R_o$, as discussed in the manuscript. Consequently, within these aggregates, the active dumbbells slow down considerably. The increased aggregation size contributes to a reduction in speeds; as a result, it leads to decay in the directed speed for larger $R_o$.
 
 Nevertheless, despite this slowdown, active dumbbells form ordered jammed structures around the circumference of the surface that generate torque on the cluster, therefore having the significant tangential speed to make them rotate around the obstacle.

Furthermore, we probe the average directed speed $\mathrm{v^r_d}(r)$ as a function of radial distance $r$ from the center of an obstacle for various $Pe$ at fixed obstacle radius $R_o = 15$.  As expected, for the passive system $Pe = 0$, the average directed speed is zero; therefore, $\mathrm{v^r_d}(r)$ remains zero at all distances,  see Fig.~\ref{fig:dir_speed_pe}-b. While for low activity ($Pe < 10$), the directed speed $\mathrm{v^r_d}(r)$  near the obstacle's surface and far from them is nearly the same due to lack of dense aggregation on the surface, in this limits the density of active dumbbells is everywhere roughly same as in bulk phase. However, for $Pe\ge 15$ aggregation of the dumbbell sets on the surface, the speed of dumbbells is significantly reduced near the surface, as illustrated in Fig.~\ref{fig:dir_speed_pe}-b. The directed speed monotonically grows as we go far from the surface until it reaches a plateau corresponding to the speed of active dumbbells in the homogeneous bulk phase. 

\begin{figure}
    \centering
    \includegraphics[width=\columnwidth]{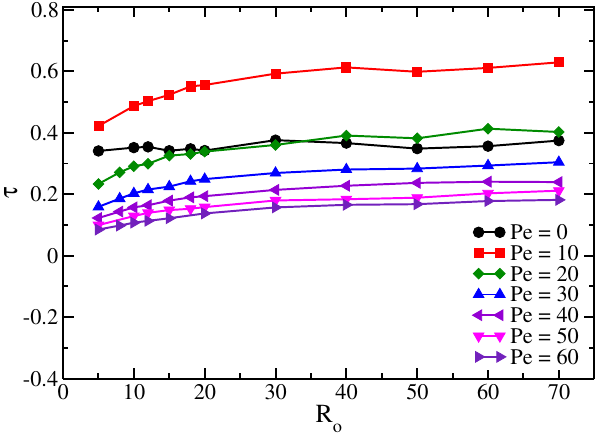}
    \caption{Variation of the residence time $\tau$ as  a function of $R_o$ for various $Pe$.}
    \label{fig:retent_time}
\end{figure}

\section{Residence time}
Now, we present the residence time of an active dumbbell on the obstacle's surface in the dilute concentration as a function of radius $R_o$ and active speed $Pe$.  We place a randomly oriented active dumbbell on the obstacle's surface to compute this time. Then, we measure the average duration a dumbbell spends within a cutoff radius of $1.5$ from the surface of a circular obstacle, called here residence time $\tau$. This calculation is averaged over 2500 randomly oriented particles on the obstacle's surface.

Figure~\ref{fig:retent_time} illustrates the variation of the residence time $\tau$ as a function of obstacle radius $R_o$ for different P\'eclet numbers $Pe$. For a passive system, i.e., $Pe = 0$,  the residence time $\tau$ remains constant across all $R_o$ values. However, for active systems, the residence time weakly increases with $R_o$ for all $Pe$.  As $Pe$ increases, the curve shifts downward, indicating that the residence time decreases for larger $Pe$. 
This reduction in residence time at higher $Pe$ values is attributed to active particles exiting in the cutoff region at higher speeds, primarily when oriented away from the surface.
In summary, our finding indicates that, in the dilute limit, the residence time of active dumbbells demonstrates only a weak dependence on both curvature radius ($R_o$) and P\'eclet number ($Pe$).  Further, the effect of the concentration of the active dumbbells leads to dramatic increases of $\tau$, as discussed in the main manuscript. 

\section{Supplementary Movies}
Here, we provide various supporting movies to illustrate the collective dynamics of active dumbbells around the static obstacle and the dynamic behavior of a tracer particle in the medium of active dumbbells. \\

Movie S1: The movie shows the dynamics of the active dumbbells at P\'eclet number  $Pe = 10$ at a given obstacle radius $R_o = 15$. Only the central portion of the simulation box is presented for clarity.\\

Movie S2: The movie shows the dynamics, aggregation, and rotation of the active dumbbells on the surface of an obstacle at P\'eclet number  $Pe = 20$ at a given radius $R_o = 15$. Only the central portion of the simulation box is presented for clarity.\\

Movie S3: The movie shows the dynamics, aggregation, and rotation of the active dumbbells on the surface of an obstacle at P\'eclet number  $Pe = 40$ at a given radius $R_o = 15$. Only the central portion of the simulation box is presented for clarity.\\

Movie S4: The movie shows the dynamics of the active dumbbells at P\'eclet number  $Pe = 40$ at a given obstacle radius $R_o = 5$. Only the central portion of the simulation box is presented for clarity..\\ 

Movie S5: The movie shows the dynamics, aggregation, and rotation of the active dumbbells on the surface of an obstacle of radius $R_o = 10$ at P\'eclet number  $Pe = 40$. Only the central portion of the simulation box is presented for clarity.\\

Movie S6: The movie illustrates the dynamic behavior of a tracer particle of radius $R_o = 15$ at  $Pe = 10$ in the solution of the active dumbbells.\\

Movie S7:  The movie illustrates the dynamic behavior of a tracer particle of radius $R_o = 15$ at  $Pe = 40$ in the solution of the active dumbbells.\\

% \textit{Introduction\textemdash}

%\section{Acknowledgements}
% \textit{Acknowledgements\textemdash} Authors acknowledge the HPC facility at IISER Bhopal for the computation time. S.P.S. thanks DST SERB Grant No. YSS/2015/000230 for the financial support. Ch.T. thanks IISER Bhopal for the funding.
% \newpage

\bibliography{reference}